\documentclass{aa}  
\usepackage{graphicx}
\usepackage{txfonts}
\usepackage{lscape}
\usepackage[colorlinks,citecolor=blue,hypertexnames=true,bookmarks=false,breaklinks=true]{hyperref}
\usepackage[dvipsnames]{xcolor}
\usepackage{makecell}
\newcommand{\acosmos}[0]{A$^3$COSMOS\xspace}
\newcommand{\agoodss}[0]{A$^3$GOODSS\xspace}

\begin{document} 

   \title{\acosmos and \agoodss: Continuum source catalogues and multi-band number counts}

   \author{Sylvia Adscheid \inst{1}
          \and
          Benjamin Magnelli \inst{2}
          \and
          Daizhong Liu \inst{3}
          \and
          Frank Bertoldi \inst{1}
          \and 
          Ivan Delvecchio \inst{4}
          \and
          Carlotta Gruppioni \inst{5}
          \and
          Eva Schinnerer \inst{6}
          \and
          Alberto Traina \inst{5,7}
          \and          
          Matthieu Béthermin\inst{8,9}
          \and 
          Athanasia Gkogkou \inst{8}          
          }

   \institute{Argelander-Institut für Astronomie, Universität Bonn, Auf dem Hügel 71, 53121 Bonn, Germany\\
              \email{sadscheid@astro.uni-bonn.de}
        \and
             Université Paris-Saclay, Université Paris Cité, CEA, CNRS, AIM, 91191, Gif-sur-Yvette, France
        \and
             Max-Planck-Institut für extraterrestrische Physik, Gießenbachstraße 1, 85748, Garching b. München, Germany
        \and
            INAF - Osservatorio Astronomico di Brera 28, 20121, Milano, Italy and Via Bianchi 46, 23807 Merate, Italy 
        \and
             Dipartimento di Fisica e Astronomia (DIFA), Università di Bologna, via Gobetti 93/2, I-40129 Bologna, Italy
        \and         
             Max Planck Institut für Astronomie, Königstuhl 17, D-69117 Heidelberg, Germany  
        \and
             Istituto Nazionale di Astrofisica (INAF) - Osservatorio di Astrofisica e Scienza dello Spazio (OAS), via Gobetti 101, I-40129
Bologna, Italy     
        \and 
            Aix-Marseille Univ., CNRS, CNES, LAM, Marseille, France
        \and
            Université de Strasbourg, CNRS, Observatoire astronomique de Strasbourg, UMR 7550, 67000 Strasbourg, France
             }

   \date{Received date; accepted date}

  \abstract
   {Galaxy submillimetre number counts are a fundamental measurement in our understanding of galaxy evolution models. 
   Most early measurements are obtained via single-dish telescopes with substantial source confusion, whereas recent interferometric observations are limited to small areas.}
   {We used a large database of ALMA continuum observations to accurately measure galaxy number counts in multiple (sub)millimetre bands, thus bridging the flux density range between single-dish surveys and deep interferometric studies.}
   {We continued the Automated Mining of the ALMA Archive in the COSMOS Field project (\acosmos) and extended it with observations from the GOODS-South field (\agoodss).
   The database consists of $\sim$4,000 pipeline-processed continuum images from the public ALMA archive, yielding 2,050 unique detected sources,   including sources   with and without a known optical counterpart. 
   To infer galaxy number counts, we constructed a method to reduce the observational bias inherent to targeted pointings that dominate the database. 
   This method comprises a combination of image selection, masking, and source weighting. 
   The effective area was calculated by accounting for inhomogeneous wavelengths, sensitivities, and resolutions and for the spatial overlap between images. 
   We tested and calibrated our method with simulations.}
   {We derived the number counts in a consistent and homogeneous way in four different ALMA bands covering a relatively large area.
   The results are consistent with number counts retrieved from the literature within the uncertainties. 
   In Band~7, at the depth of the inferred number counts, $\sim$40\% of the cosmic infrared background is resolved into discrete sources. This fraction, however, decreases with increasing wavelength, reaching $\sim$4\% in Band~3.
   Finally, we used the number counts to test models of dusty galaxy evolution, and find a good agreement within the uncertainties.}
   {By continuing the \acosmos and \agoodss archival effort, we obtained the deepest archive-based (sub)millimetre number counts measured to date over such a wide area. This database proves to be a valuable resource that, thanks to its substantial size, can be used for statistical analyses after having applied certain conservative restrictions.}


   \maketitle

\section{Introduction}

A central aspect in the study of galaxy evolution is the understanding of the cosmic star formation history. The star formation rate (SFR) of galaxies is commonly traced by emission in the rest-frame ultraviolet (UV) and/or far-infrared (far-IR). 
The presence of short-lived massive stars emitting strongly at UV wavelengths indicates recent star formation activity;  part of this UV radiation is absorbed by dust and re-emitted in thermal IR. 
Galaxy studies in UV to IR wavelengths have shown that the cosmic SFR density was much higher in the past ($\sim$10$\times$ higher at its peak at $z\approx2$), with most of star formation activity being obscured by dust \citep[$\sim$80\%; see][for a review]{madau14}. 
Dusty star forming galaxies (DSFGs) contribute significantly to the cosmic SFR density; they are bright in the submillimetre regime and faint or even undetected in rest-frame UV to optical \citep[e.g.][]{casey14,wang19,talia21}.
The number and flux density distribution of DSFGs results from the complex history of gas accretion, star formation, and dust production and their underlying physical mechanisms. 
The number density of galaxies above a given flux density threshold per unit effective area, commonly referred to as number count, is therefore a fairly simple measure, but a very useful tool for constraining galaxy evolution models. 
A plethora of single-dish surveys at wavelengths $\geq450\,\mu$m have been conducted over the last two decades to reveal the nature and properties of DSFGs, for example the SCUBA Lens Survey \citep{smail02}, the LESS survey \citep{weiss09}, the Lockman Hole North 1.2\,mm survey \citep{lindner11}, six 1.1\,mm AzTEC blank-fields \citep{scott12}, the SCUBA-2 COSMOS survey \citep{casey13}, the ACT Southern survey \citep{marsden14}, the GISMO 2\,mm survey \citep{magnelli19}, S2COSMOS \citep{simpson19}, STUDIES \citep{wang17,dudzeviciute21}, and the N2CLS survey \citep{bing23}. 
These single-dish surveys have already set valuable constraints on the  (sub)millimetre number counts. 
However, due to their large beam sizes, they are limited by source confusion to the brightest DSFGs. 
This leaves a large part of the population of DSFGs and the lower flux density regimes of models largely unconstrained \citep[e.g.][]{bethermin12,bethermin17,casey18,popping20}. 

The  Atacama Large Millimeter/submillimeter Array (ALMA) can help to alleviate these issues, being one of the highest-sensitivity instruments currently operating in the (sub)millimetre regime; it achieves resolutions that far exceed the capabilities of single-dish telescopes at these wavelengths due to its interferometric nature.
However, due to its small field of view, deep blind surveys with ALMA  are very time-consuming and hardly viable for very extended fields. 
Though some blind surveys have been performed with ALMA, such as ASAGAO \citep{hatsukade18}, ASPECS-LP \citep{gonzales-lopez19,gonzales-lopez20}, MORA \citep{zavala21,casey21}, and the GOODS-ALMA survey \citep{franco18,gomez-guijarro21}, they are small in size compared to single-dish surveys (e.g. $\sim$70\,arcmin$^2$ for GOODS-ALMA vs $\sim$2\,deg$^2$ for S2COSMOS). 
Instead, ALMA projects are more often designed as follow-ups to galaxy samples from larger single-dish programmes \citep[e.g.][]{karim13,stach19,simpson20}. 
Therefore, they are focussed on the brighter, more starbursty galaxies, neglecting the bulk of the DSFG population.  

The aim of the \acosmos project\footnote{\url{https://sites.google.com/view/a3cosmos}} \citep[][]{a3cosmos_1,a3cosmos_2} is to aggregate and homogeneously process all public ALMA archival data in the $\sim$2\,deg$^2$ Cosmic Evolution Survey \citep[COSMOS;][]{scoville07} in order to provide to the community homogeneously processed images\footnote{\url{https://irsa.ipac.caltech.edu/data/COSMOS/images/a3cosmos/}} and catalogues of prior-based and blind (sub)millimetre photometry and coherently derived galaxy properties. 
Among other blind deep fields (e.g. GOODS-North/South, UDS, EGS), COSMOS has the largest \textit{HST}/ACS contiguous coverage \citep{scoville07} and  the largest \textit{JWST}/NIRCam (0.54\,deg$^2$) and MIRI (0.19\,deg$^2$) coverage \citep[COSMOS-Web;][]{casey22}.
Due to this rich multi-wavelength coverage and the legacy status of COSMOS, the ALMA coverage in COSMOS, and therefore the \acosmos database, is extensive and continuously growing. 
The \acosmos database is naturally dominated by single pointings and their respective targets, but these pointings are usually far more extended than the angular size of their targets, which also yields an abundance of serendipitous detections.  
This facilitates the study of galaxy number counts utilising the large sky coverage of \acosmos.  

In this paper, we present the latest data version of \acosmos, for which we included another well-studied extragalactic field:
the Great Observatories Origins Deep Survey Southern field \citep[GOODS-S;][]{dickinson03}.
Although the areal size of this field is smaller than COSMOS, originally covering $\sim$160\,arcmin$^2$, it offers rich ancillary data from a number of deep survey programmes, for instance hosting the Hubble Ultra Deep Field \citep[HUDF;][]{hudf}. 
The ancillary data available in GOODS-S is generally deeper than in COSMOS, which facilitates the study of fainter and higher-redshift galaxies.
We then use the combined database to homogeneously derive galaxy number counts in several ALMA bands. 

The structure of this paper is as follows.
In Sect. \ref{sec_a3cosmos} we describe the updated \acosmos database and additions made to the processing pipeline, including the GOODS-S field (i.e. \agoodss). 
In Sect. \ref{sec_nc} we use the combined database to infer number counts in multiple ALMA bands. 
We describe the calculation of the effective area and corrections made to reduce the observational biases.
These corrections are then tested using simulations. 
The results are discussed in Sect. \ref{sec_results}, and summarised in Sect. \ref{sec_summary}.

In the following we assume a flat $\Lambda$CDM cosmology with $H_0 = 70$\,km\,s$^{-1}$\,Mpc$^{-1}$, $\Omega_{\Lambda}$ = 0.7, $\Omega_M = 0.3$, and a \citet{chabrier03} initial mass function.

\section{The \acosmos and \agoodss database} \label{sec_a3cosmos}

The goal of \acosmos is to assemble a large sample of galaxies detected at (sub)millimetre wavelengths and probing a wide redshift range ($z \approx 0-6$). This is done by using a pipeline to retrieve publicly available observational data from the ALMA archive and homogeneously processing them into continuum images from which sources are then extracted. Galaxy properties (e.g. SFR, stellar mass, dust luminosity) are inferred by fitting the spectral energy distributions (SEDs) of the extracted galaxies, making use of both the extracted ALMA flux densities and ancillary data (see Sect. \ref{subsec_cosmos} and \ref{subsec_goodss}). 

For the new data version (\texttt{20220606}), we extended the database to ALMA observations in GOODS-S. 
This not only improves the statistical properties of our sample by including more galaxies, but it also offers the opportunity to expand our sample to fainter objects due to the availability of deeper ancillary photometric and redshift information, even though the areal size of the GOODS-S field is much smaller than that of COSMOS.

\subsection{The \acosmos pipeline} \label{subsec_a3cosmos_pipeline}

For a full in-detail description of the entire pipeline of data retrieval, processing, source extraction and SED fitting, we refer the reader to the original work by \citet{a3cosmos_1}. 
Here, we give a brief summary of the most important steps.
The photometric data used in the pipeline, as well as any updates and additions to those with respect to the previous data version, are listed separately in Sects. \ref{subsec_cosmos} and \ref{subsec_goodss}.

The archive query and download were done via the Python package \texttt{astroquery} \citep[][]{astroquery}.
Calibration and creation of continuum images from the raw data was done using the  Common Astronomy Software Application \citep[CASA;][]{casa} in the recommended version for each individual ALMA project. 
For calibration, the \texttt{scriptForPI.py} provided by the ALMA Observatory was used. 
Imaging was performed in automatic pipeline mode, using Briggs-weighting with \texttt{robust = 2.0}, and choosing \texttt{specmode = `cont'}. 
The continuum images were masked outside of a primary beam attenuation of 0.2.
Source extraction from the produced images was then done in two separate ways: 
`blind' extraction was performed using the Python Blob Detector and Source Finder \citep[\texttt{PyBDSF};][]{pybdsf}; `prior' extraction was performed with our \acosmos prior extraction pipeline \citep[see Sect. 2.4 in][]{a3cosmos_1} that iteratively executes \texttt{GALFIT} \citep[][]{galfit1, galfit2} using positional priors of known sources from a previously assembled master catalogue.
Sources are fit simultaneously in \texttt{GALFIT} to allow for a dissection of sources potentially blended in blind extraction. 
Fortunately, due to the high resolution of most ALMA observations, source blending plays only a minor role, affecting only $\sim$$2.3\%$ of the sources in our blind catalogue, most of which are the targets of their respective observation.
These blended sources are indicated by a flag in our blind source catalogue.
We note that there are no blended sources in the source sample of our number counts analysis in Sect.~\ref{sec_nc}.

The master catalogue used for prior extraction is a compilation of multiple multi-wavelength catalogues covering the COSMOS and GOODS-S fields (see Sect. \ref{subsec_cosmos} and \ref{subsec_goodss}, respectively), spatially cross-matched with a matching radius of 1\,arcsec to ensure that all entries are unique sources. 
This radius corresponds to a false-match probability of 13.3\% between the COSMOS2015 catalogue from \cite{cosmos2015} and other catalogues \citep[see Sect. 2.3 in][]{a3cosmos_1}.
To avoid a high contamination with spurious detections in both source extraction meth\-ods, a minimum peak signal-to-noise (S/N$_{\rm peak}$) threshold was applied. 
This threshold was set to 5.40 for \texttt{PyBDSF} (blind) and 4.35 for \texttt{GALFIT} (prior) extraction, as to limit the cumulative spurious fraction to $\lesssim$8\% and $\lesssim$12\% \citep[see Sect. 2.8 and 3.2 in][]{a3cosmos_1}. 
Sources with S/N$_{\rm peak}$ below these respective thresholds were discarded for further analysis (but kept in our released photometry catalogues)\footnote{These catalogues also provide information on the respective image RMS noise and beam size of each (non-)detection, which users can utilise to infer upper limits on the flux densities of sources below the detection threshold in the prior catalogue (see also Sect. \ref{subsubsec_read_map}).}.
The blind and prior source catalogues were then spatially cross-matched, with a matching radius of 1\,arcsec, to create a combined sample of unique sources, retaining the information from both prior and blind extraction for sources retrieved through both methods. 
The ALMA extracted conti\-nu\-um flux densities of the sources in this combined sample were then combined with the available ancillary photometry and photometric and/or spectroscopic redshift information of their respective priors (naturally, except for sources unique to the blind catalogue). 

To assess the reliability of the master catalogue priors as counterparts for the extracted ALMA sources (which we call counterpart association or `CPA'), we defined a number of parameters, such as total flux density signal-to-noise, spatial separation and source extension, and measured them at the prior and extracted source coordinates in both the ALMA images and different optical and near-infrared counterpart images (listed in Sects. \ref{subsec_cosmos} and \ref{subsec_goodss}). 
Based on these parameters, counterparts were classified as reliable or unreliable using a combination of visual inspection and machine learning.
This process is described in detail in Sect. 4.2 of \citet{a3cosmos_1} and summarised in Appendix \ref{appendix_cpa}.
Sources with unreliable counterparts, called `CPA discarded', were excluded from further analysis ($\sim$9\% in total, see Table \ref{tab_database}). 

To infer physical galaxy properties from the sample of ALMA sources with reliable optical-to-near-IR counterparts, we used the \texttt{MAGPHYS} SED fitting package with its high-redshift library update \citep[][]{magphys,magphys2}, which is applicable to both low- and high-redshift galaxies. 
This package finds the best fit SED by comparing the given photometry with a comprehensive library of SED models via a reduced $\chi^2$ approach. 
If several redshifts were available in our master catalogue for one ALMA source, the one yielding the lowest $\chi^2$ was chosen. 
In the process, the potential contamination of the ALMA continuum flux densities from line emission was assessed and subtracted using the information on the spectral set-up of the ALMA observations and empirical IR-to-line luminosity correlations for the brightest (sub)millimetre lines, that is, [\textsc{Ci}] \citep{liu15}, [\textsc{Cii}] \citep{delooze11}, [\textsc{Nii}] \citep{zhao16}, and CO \citep{sargent14,liu15}.
Since \texttt{MAGPHYS} does not include an AGN component, strong mid-IR contribution from AGNs can cause an overestimation of inferred stellar masses and SFRs. 
\citet{a3cosmos_1} showed that by running \texttt{MAGPHYS} iteratively, once with and once without considering any 24\,$\mu m$ flux density information, and choosing the fit yielding the lowest $\chi^2$, the inferred stellar masses are in a good agreement with results from \citet{delvecchio17} obtained from SED fitting including an AGN component \citep[see Sect. 4.6 in][]{a3cosmos_1}.
Therefore, \texttt{MAGPHYS} was run twice, one time disregarding any potential mid-IR flux density information from the \textit{Spitzer}/MIPS 24\,$\mu m$ filter, and the better fit was chosen.
The flux densities at all wavelengths of the extracted ALMA sources ($S_{\rm OBS}$) were compared to those inferred from the best fit SEDs ($S_{\rm SED}$) by plotting the distribution of the observed-to-predicted flux density ratios for all sources. Fitting a 1D Gaussian function to this distribution, we found a mean of $\mu=1$ and a scatter of $\sigma\approx0.05$\,dex. 
Then, sources were discarded if their observed-to-predicted flux density ratio at any wavelength differed by more than 5\,$\sigma$ ($\sim$0.25\,dex) from a ratio of one \citep[called `SED outliers', see Sect.~4.4 in][also \citealt{traina23}]{a3cosmos_1}. 
This deviation likely indicates a mismatch between the measured ALMA flux densities and the ancillary photometry, hence implying a chance association of sources ($\sim$7\% of sources, see Table~\ref{tab_database}).
We note that sources can be classified simultaneously as both CPA discarded and SED outliers, which, however, does not apply to all cases.
We also investigated the use of the distribution in $(S_{\rm OBS}-S_{\rm SED})/\sigma_{\rm OBS}$ as a measure of fit accuracy, which incorporates the uncertainty $\sigma_{\rm OBS}$ on the observed flux density.  
However, in addition to the uncertainty on the observed photometry, the quality of our fits is also determined by other factors, such as uncertainties from the ancillary photometry, fluctuations in ALMA as well as ancillary zero-point calibration, and \texttt{MAGPHYS} fitting limitations.
Through a visual assessment, we found the simple flux ratio to overall yield the best balance between all those factors and thus to be the best method to identify obvious outliers.

From the best-fit SED, we obtained galaxy properties, such as stellar mass, dust mass, SFRs averaged over the last 100\,Myr of their star-formation history, dust temperature, and total infrared luminosity. 
From the total infrared luminosity of these galaxies, we also inferred empirically derived SFRs using the calibration from \citet{kennicutt98} and a \citet{chabrier03} initial mass function.
Finally, molecular gas masses were inferred using the rest-frame 850\,$\mu$m flux density of these galaxies and the Rayleigh-Jeans continuum method applying the calibration of \citet{hughes17}. 
This calibration was chosen in accordance with \citet{a3cosmos_2}, as it uses a luminosity-dependent conversion factor, as opposed to the commonly used calibration from \citet{scoville17} which uses a single conversion factor. 
We note, however, that the \citet{hughes17} calibration yields systematically lower molecular gas masses (by $\sim$0.1\,dex) than the one from \citet{scoville17}, as found in a direct comparison carried out by \citet[][see their Sect. 3]{a3cosmos_2}.

We note that some ALMA projects contain several coverages of the same pointing or mosaic. 
While combining the individual images in the \textit{uv}-plane would result in a single deeper continuum image, we chose to instead pipeline-process all images homogeneously and thus separately. 
Hence, one source can have several flux density estimates at the same wavelength. 
This has no importance for our SED fit, as both information are mathematically combined via the $\chi^2$. 
But it also means that sources lying slightly below the noise thresholds in those images could possibly have been detected in the combined image. 
This is a current limitation of the pipeline, which we plan to improve in future versions.
Fortunately, for some wide and/or deep surveys in the COSMOS and GOODS-S fields, the already combined continuum maps are publicly available.
Performing source extraction on these maps (instead of the individual coverages from our pipeline) allows us to circumvent the depth limitation, at least in these specific surveys, and thus extend our galaxy number counts to lower flux densities.
This is detailed in Sect. \ref{subsec_comb_mosaics}.

In the following, we describe the update of the database, the changes in ancillary data with respect to the previous data release, and introduce the new extension to the GOODS-S field, \agoodss.

\subsection{COSMOS} \label{subsec_cosmos}

\begin{figure}
\centering
\includegraphics[width=\linewidth]{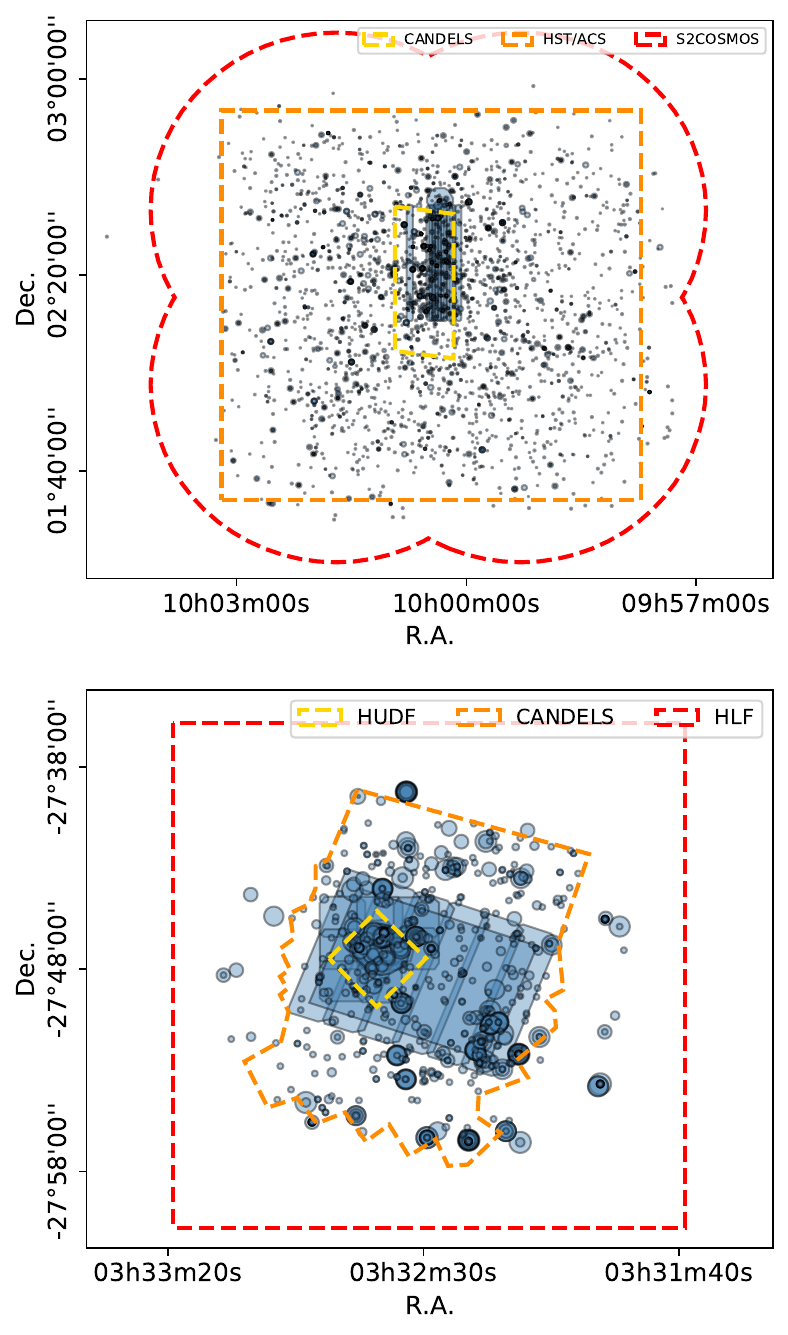}
\caption{Spatial sky coverage of all successfully imaged ALMA archival data available for COSMOS (\textit{top panel}) and GOODS-S (\textit{bottom panel}) as of June 6, 2022. The blue shaded shapes are outlines of single pointings and blind surveys; the darker shades of blue indicate an overlap of multiple images. For single pointings, the circle size corresponds to the area where the primary beam attenuation is >0.5. The outlines of blind surveys have been traced manually. For easier visibility the outlines of surveys are only drawn once, even if multiple coverages of the same survey exist in our database. The dashed coloured lines indicate the approximate outlines of different ancillary survey fields. In COSMOS: CANDELS \citep[yellow;][]{grogin11}, COSMOS \textit{HST}/ACS imaging \citep[orange;][]{koekemoer07}, and S2COSMOS \citep[red;][]{simpson19}. In GOODS-S: Hubble Ultra Deep Field \citep[yellow;][]{hudf}, CANDELS \citep[orange; as in][]{guo13}, and Hubble Legacy Fields \citep[red;][]{hlf}.}
\label{sky_coverage}
\end{figure}

We downloaded and pipeline-processed all ALMA archive data that became public up until June 6, 2022, as opposed to January 2, 2018 \citep[version number \texttt{20180102};][]{a3cosmos_1}.  
The new images were added to the database, extending it from $\sim$1,500 continuum images in \citet{a3cosmos_1} to $\sim$3,300 from 173 individual ALMA projects. 
The spatial coverage of all produced continuum images is displayed in the top panel of Fig.~\ref{sky_coverage}.
We also show the coverage in each individual ALMA band in Fig.~\ref{a3cosmos_coverage_by_band}.
Most pointings are located within the area of the COSMOS \textit{HST}/ACS field \citep[][]{koekemoer07}, covering it with nearly uniform density. 
We notice the existence of large mosaics from two blind surveys: MORA in Band~4 \citep[][vertical stripes]{zavala21, casey21} and a hexagonal field in Band~3 \citep[][]{keating20}.

The master catalogue was also updated. 
The individual prior catalogues used to assemble the previous master catalogue are listed in Table~2 of \citet{a3cosmos_1}. 
We replaced the COSMOS2015 catalogue \citep{cosmos2015} by the newer version COSMOS2020 \citep{cosmos2020}. 
Two catalogues are available for COSMOS2020: 
the \textsc{Classic} catalogue and the \textsc{Farmer} catalogue, which use two different photometry methods. 
Since the two catalogues differ slightly in their spatial sky coverage, we included both. 
Additionally, two more prior catalogues were included: 
the \textit{Spitzer}/IRAC catalogue from \citet{ashby18} and the \textit{Spitzer}/24$\,\mu$m catalogue from \citet{lefloch09}, which respectively add another 85,466 and 804 priors not contained in any of the other catalogues. 
With this updated master catalogue, we were able to identify a counterpart for $\sim$93\% of our extracted ALMA sources in COSMOS (see Table~\ref{tab_database}), which is a notable improvement over the previous data version \citep[$\sim$74\% in][]{a3cosmos_1}.

The ancillary photometry of each ALMA source was taken from each of the catalogues that constitute the master catalogue, complemented by additional deblended mid- to far-IR photometry from \citet{jin18}. 
Photometric redshifts were taken from the COSMOS2020 catalogue \citep{cosmos2020}, \citet{salvato11}, \citet{davidzon17}, \citet{delvecchio17}, and \citet{jin18}. 
Spectroscopic redshifts were adopted from the compilation by M. Salvato as listed in Sect.~2 of \citet{a3cosmos_1}, as well as supplementary spectroscopic redshifts from \citet{riechers14}, \citet{capak15}, \citet{smolcic15}, \citet{brisbin17}, \citet{lee17}, and \citet{pavesi18}.

For our CPA (see Sect.~\ref{subsec_a3cosmos_pipeline} and Appendix~\ref{appendix_cpa}), we made use of the following counterpart images:  
\textit{HST}/ACS i-band from COSMOS \citep[][]{koekemoer07,massey10}, \textit{Spitzer}/IRAC at 3.6\,$\mu$m from the Cosmic Dawn survey \citep[][]{moneti22}, VISTA/VIRCAM K$_\textup{s}$-band from the UltraVISTA survey \citep[release version DR4;][]{mccracken12} and VLA at 3\,GHz from the VLA-COSMOS 3\,GHz Large Project \citep[][]{smolcic17}.

\subsection{GOODS-South} \label{subsec_goodss}

We downloaded all public ALMA archive data available in the GOODS-S field at the same time as COSMOS, that is, on June 06, 2022. 
The search was centred on the position $\textup{R.A.}=53.125$° and $\textup{Dec.}=-27.806$°, with a radius of 10\,arcmin. 
Data calibration and continuum imaging was performed using the pipeline as described in Sect.~\ref{subsec_a3cosmos_pipeline}, yielding $\sim$700 images from 74 ALMA projects. 
The total spatial coverage is shown in the bottom panel of Fig.~\ref{sky_coverage}, and the individual coverage in each ALMA band in Fig.~\ref{a3goodss_coverage_by_band}.
The pointings are mostly located inside the CANDELS field \citep[][]{guo13}, but with the area within the HUDF being sampled far more densely than the rest of the field. 
A number of blind surveys are available, most importantly the GOODS-ALMA survey in Band~6 \citep{franco18,gomez-guijarro21}, the ASAGAO survey in Band 6 \citep{hatsukade18}, the ASPECS-LP survey in Bands~3 and 6 \citep{decarli19,gonzales-lopez19, gonzales-lopez20} and a wide rectangular field in Band~3 (PI: R. Decarli). 

We assembled a new master catalogue for the GOODS-S field and its surrounding area, using positional priors and photometry in UV to IR wavelengths from the CANDELS GOODS-South catalogue from \citet{guo13}, the 3D-HST/CANDELS programmes in GOODS-S from \citet{skelton14}, the S-CANDELS survey in CDFS \citep{ashby15}, the ZFOURGE survey in CDFS \citep{straatman16} and in radio wavelength from the VLA 1.4\,GHz survey in E-CDFS by \citet{miller08}. 
The final master catalogue contains 82,519 priors over an area of $\sim$1,300 arcmin$^2$, with the majority of priors concentrated in a region of $\sim$650 arcmin$^2$ roughly centred on the HUDF. 
We also included additional far-IR photometry from the PEP-GOODS-\textit{Herschel} programme \citep{magnelli13}. 

We adopted photometric redshifts from \citet{skelton14}, \citet{straatman16}, \citet{croom01}, \citet{wuyts08}, and \citet{momcheva16}. 
Spectroscopic redshifts were taken from \citet{lefevre04}, \citet{mignoli05}, \citet{ravikumar07}, \citet{vanzella08}, \citet{balestra10}, \citet{lefevre13}, \citet{kriek15}, \citet{morris15}, \citet{tasca17}, \citet{aravena19}, and \citet{urrutia19}. 

For our CPA, we adopted counterpart images from: 
\textit{Spitzer}/IRAC at 3.6 µm and 4.5 µm from the SEDS survey \citep{ashby13}, \textit{HST}/ACS F775W-band from the GOODS survey \citep{giavalisco04}, \textit{HST}/WFC3 F160W-band from the CANDELS survey \citep{grogin11, koekemoer11,skelton14}, and CFHT/WIRCam K$_\textup{s}$-band from the TENIS survey \citep{hsieh12}.

\subsection{Adding combined survey mosaics} \label{subsec_comb_mosaics}

Combined continuum maps are available for the following ALMA surveys:
GOODS-ALMA\footnote{private communication} \citep[at 1.1\,mm;][]{gomez-guijarro21}, ASPECS-LP\footnote{\url{https://almascience.nrao.edu/almadata/lp/ASPECS/}\label{footnote_aspecs}} \citep[at 1.2\,mm and 3\,mm;][]{gonzales-lopez19,gonzales-lopez20}, and MORA\footnote{\url{https://www.as.utexas.edu/~cmcasey/alma2mmcosmos/}} \citep[at 2.1\,mm;][]{casey21}.
We replaced our pipeline-imaged maps of these surveys by these publicly available combined maps to allow for a deeper detection limit (see Sect.~\ref{subsec_a3cosmos_pipeline}). 
However, unlike our pipeline-processed images, these combined maps were not `cleaned’, hence their point spread function (PSF) differed from an ideal 2D Gaussian function (dirty vs clean beam). 
This means that while \texttt{PyBDSF} could be used to detect sources in these maps, it could not be used to accurately measure their flux densities. 
We therefore had to resort to another approach.

For GOODS-ALMA and ASPECS-LP, the PSFs are available alongside the continuum maps, which allows for an aperture photometry approach:
integrated flux densities were measured within circular apertures placed at the positions of source candidates (i.e. sources for which \texttt{PyBDSF} yields a S/N$_{\rm peak}\geq5.4$) and normalised by the PSF.
We verified that the so-extracted flux densities were in agreement with measurements from the original works of \citet{gomez-guijarro21} and \citet{gonzales-lopez19,gonzales-lopez20}. 
Naturally, this approach does not yield any information about the size of the sources, information that is therefore absent from our blind catalogue. 
While we do not provide modelled source size information in our blind source catalogue, an estimate of the angular extent of sources in GOODS-ALMA and ASPECS-LP can, however, still be deduced from the ratio of their peak to integrated flux densities.
We used this approach for the computation of our number counts, which requires source size information (see Sect.~\ref{subsec_effective_area}).
We note, however, that sources are mainly considered as point source at the coarse angular resolution of the ASPECS-LP survey ($1.5\times1.1$\,arcsec and $1.8\times1.5$\,arcsec at 1.2\,mm and 3\,mm, respectively) and only marginally resolved at the intermediate angular resolution of the GOODS-ALMA survey ($0.7\times0.5$\,arcsec). 
This is consistent with the finding from the original works of \citet{gonzales-lopez19,gonzales-lopez20} and \citet{gomez-guijarro21}.

For the MORA map, the PSF is not publicly available, preventing us from measuring flux densities via our aperture photometry approach.
Fortunately, for this survey the flux densities obtained by \texttt{PyBSDF} assuming a 2D Gaussian PSF were in very good agreement with those from the original work of \citet{casey21}. 
We only noted a slight overestimation by 10\% of our \texttt{PyBDSF} flux densities, which we simply corrected by multiplying our flux densities by a factor 0.9.
Finally, in line with \citet{casey21} and with the very coarse angular resolution of this survey ($1.8\times1.4$\,arcsec), we treated these detections as point sources.

\subsection{Summary of \acosmos and \agoodss data release \texttt{20220606}}

\begin{table}
  \caption[]{Properties of the \acosmos and \agoodss databases.}
     \label{tab_database}
 $$ 
     \begin{tabular}{lrr}
        \hline
        \noalign{\smallskip}
           &  COSMOS  & GOODS-S \\
                &   &      \\ 

        \noalign{\smallskip}
             Database  &   &      \\         
        \hline
        \noalign{\smallskip}
             ALMA projects         & 173   &   74    \\ 
             Continuum images      & 3232  &  723    \\
             Blind survey images   & 14    &   49    \\                
        \noalign{\smallskip}
        \hline
                &   &      \\ 

        \noalign{\smallskip}
             Spatial coverage [arcmin$^2$]  &   &      \\         
        \hline
        \noalign{\smallskip}
             Band 3   & 427.7 (205.6)   &  101.0 (77.4)       \\ 
             Band 4   & 224.8 (205.2)   &   44.5 (21.2)       \\ 
             Band 5   &   2.6   (1.1)   &    1.9  (1.0)       \\ 
             Band 6   & 289.3 (127.9)   &   90.7 (86.8)       \\ 
             Band 7   & 183.3  (81.0)   &   38.5 (17.8)       \\ 
             Band 8   &   3.3   (1.4)   &    0.1  (0.05)       \\ 
             Band 9   &   0.08  (0.04)  &    0.5  (0.2)       \\
        \noalign{\smallskip}            
        \hline            
        \noalign{\smallskip}             
             Combined    & 896.3 (519.6)  &  149.6 (115.9)     \\                 
        \noalign{\smallskip}
        \hline
                &   &      \\ 
                
        \noalign{\smallskip}
             Photometry catalogues   &   &      \\         
        \hline
        \noalign{\smallskip}
         Blind catalogue   &  2204   &  491    \\
         Prior catalogue   &  2540   &  568    \\ 
        \noalign{\smallskip}
        \hline
                &   &      \\ 
        \noalign{\smallskip}
             Unique sources   &   &      \\ 
        \hline
        \noalign{\smallskip}   
            Total                    & 1756  &  294  \\  
            Blind-only               &  131  &   79  \\
            With prior photometry    & 1625  &  215  \\  
            With prior redshift      & 1514  &  185  \\   
            SED outliers             &   86  &  37   \\   
            CPA discarded            &  117  &  60   \\
        \noalign{\smallskip}            
        \hline            
        \noalign{\smallskip}
            Robust Galaxy Catalogue    & 1335  &  127    \\              
        \noalign{\smallskip}
        \hline
     \end{tabular}
 $$ 
Notes: Listed by field are the number of ALMA projects, pipeline-produced continuum images, and blind surveys; spatial coverage (taking overlapping areas into account) per ALMA band considering the area where the primary beam attenuation is >0.2 (in brackets: >0.5, as displayed in Fig. \ref{sky_coverage}); number of valid detections in the blind (S/N$_{\rm peak}$>5.4) and prior (S/N$_{\rm peak}$>4.35) catalogues; number of unique sources with prior photometry available and with prior photometric or spectroscopic redshift information (both required for SED fitting), sources discarded due to disagreement with the SED fit (SED outliers), sources discarded due to unreliable counterpart association (CPA discarded), and final number of sources in our Robust Galaxy Catalogue including galaxy properties inferred from SED fitting.
 
\end{table}

Data products from this new data version of the \acosmos project, containing both the \acosmos and \agoodss databases, are publicly available at the CDS\footnote{\label{footnote_cds}\url{https://cdsarc.cds.unistra.fr/viz-bin/cat/J/A+A/685/A1}} (for brevity, in the following we refer to \acosmos and \agoodss jointly as `\acosmos database'). 
This includes the blind and prior photometry catalogues, which contain all individual ALMA detections, and the Robust Galaxy Catalogue, which contains our final sample of unique galaxies (i.e. without CPA discarded and SED outlier sources) with their corresponding galaxy properties. 
For the description of the columns in those catalogues we refer to Tables~4 and 5 of \citet{a3cosmos_1}.
The properties of our final database and source catalogues are listed in Table~\ref{tab_database}. 
The ALMA bands with the most observations available are 3, 4, 6, and 7, each with a total spatial coverage of over 200\,arcmin$^2$ combining both COSMOS and GOODS-S. 
Band~6 profits especially from the addition of the \agoodss database, expanding the total sky area covered in this band by $\sim$30\% compared to the coverage in COSMOS, due to the availability of large blind surveys. 
In Bands~3, 4, and 7 the total coverage is increased by $\sim20-24\%$ compared to COSMOS. 
Band~5, 8, and 9 are much less commonly used and contribute only a small number of single pointings each. 

\begin{figure}
\centering
\includegraphics[width=\linewidth]{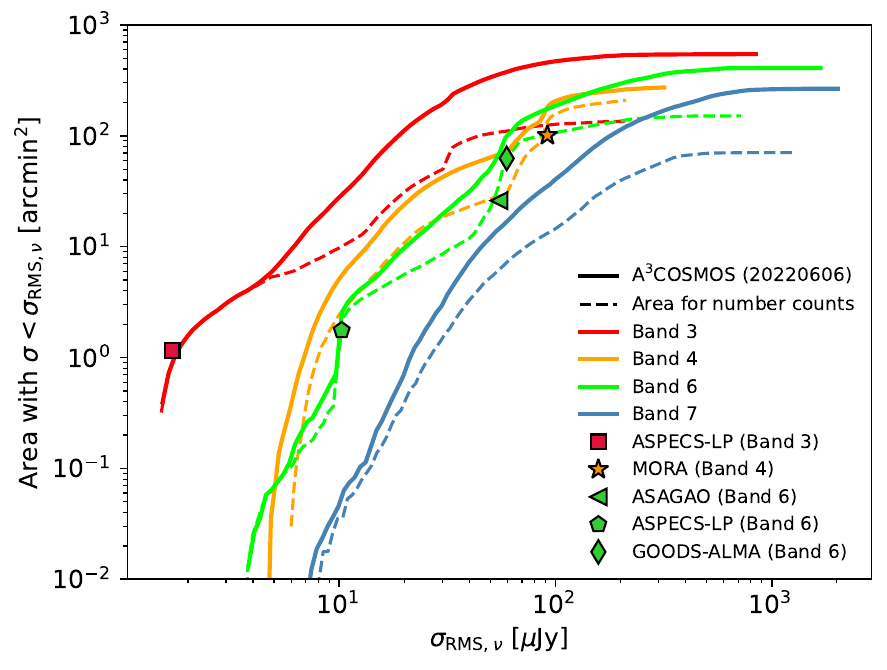}
\caption{Cumulative areal coverage of the \acosmos database, version \texttt{20220606}, as a function of 1\,$\sigma_{\rm RMS}$ sensitivity for ALMA Bands 3, 4, 6, and 7. The noise of all images in each band is normalised to the central wavelength of the band, and spatial overlaps of images are accounted for. The solid lines indicate an  entire database; the image areas are considered out to a primary beam attenuation of 0.2. The  dashed lines are images selected for the computation of the number counts after masking out the phase centres and cropping the edges at a primary beam attenuation of 0.3 (see Sect. \ref{subsec_selection_criteria}). The symbols (see legend) represent the area and normalised depth of multiple ALMA blind surveys: ASPECS-LP \citep{gonzales-lopez19, gonzales-lopez20}, MORA \citep{zavala21,casey21}, ASAGAO \citep{hatsukade18}, and GOODS-ALMA \citep{gomez-guijarro21}. }
\label{sensitivity_curve}
\end{figure}

The areal coverage as a function of depth (i.e. 1\,$\sigma$ pixel RMS noise) is shown in Fig. \ref{sensitivity_curve}.
Single-dish surveys routinely cover areas of several hundred square arcminutes down to typical sensitivities of the order of $\sim$1\,mJy.
At this depth, our database covers $\sim$200-500\,arcmin$^2$ depending on the ALMA band, approaching the regime of single-dish coverage but without their inherent source confusion problem.
Deep interferometric surveys achieve much better sensitivities ($\sim$0.01-0.1\,mJy depending on wavelength), but cover only small fields up to a few tens of square arcminutes. 
Se\-ve\-ral of these deep surveys are included in our database, plus a number of even deeper single pointings. 
Finally, the large number of single pointings of varying sensitivity yield a continuous increase in area towards shallower depths, thus bridging the regimes of deep interferometric and large single-dish surveys.

\begin{figure}
\centering
\includegraphics[width=\linewidth]{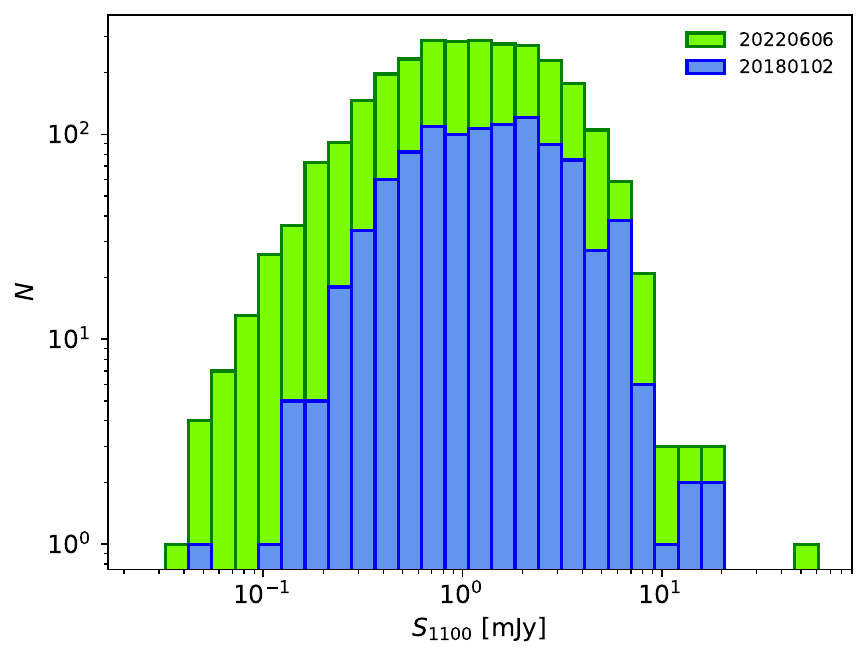}
\caption{Flux density distribution of detected sources in ALMA Bands 6 and 7 in the \acosmos database versions \texttt{20180102} and \texttt{20220606} (including \agoodss). Flux densities are normalised to 1100\,$\mu$m assuming that they probe the Rayleigh-Jeans tail of the dust continuum, with a dust spectral index $\beta=1.8$ \citep{scoville14}.}
\label{comp_old_new_dataversion}
\end{figure}

With the new data version we more than double the number of detections compared to the last release.
Figure~\ref{comp_old_new_dataversion} shows the normalised flux density distribution of unique ALMA detections in Bands~6 and 7 for the old release \texttt{20180102} and our new data version combining both \acosmos and \agoodss sources.
The largest increase in source numbers is in the intermediate to lower flux density range (i.e. <2\,mJy).
The intermediate flux density regime (i.e. 0.2-2\,mJy) benefits in particular from a large increase, in accordance with the fact that it is the most densely sampled regime in the first place.
The lower flux density regime (i.e. <0.2\,mJy), however, has the largest percentage increase with approximately an order of magnitude compared to the previous release.

Finally, we note that the total number of unique sources in the \agoodss catalogue is significantly lower than the number in the \acosmos catalogue, but this difference is understandable when considering the difference in spatial coverage (i.e.  149.6\,arcmin$^2$ vs 896.3\,arcmin$^2$). 
In fact, we verified that the number density of ALMA detections in the GOODS-S field per unit area at a given flux density (i.e. the number counts, see Sect.~\ref{sec_nc}) matches that in the COSMOS field. 
However, the fraction of unique sources entering our final Robust Galaxy Catalogue is lower in GOODS-S ($\sim$43\%) than in COSMOS ($\sim$76\%). 
There are two reasons for this difference.
Firstly, in GOODS-S there is a higher fraction of sources without any optical/IR counterpart.
This can be explained by deeper ALMA observations in GOODS-S and by the presence of more deep blind surveys in this database, which are tailored to detect sources without previously known optical counterparts. 
And secondly, GOODS-S also has a higher fraction of discarded sources due to unreliable counterparts, identified either through our dedicated CPA approach (i.e. CPA discarded) or revealed by bad SED fits (i.e. SED outliers). 
The ancillary data available in GOODS-S is generally deeper than in COSMOS, which increases the surface density of potential counterparts and thus the ambiguity of our CPA.

\section{Selection bias correction towards accurate number counts} \label{sec_nc}

\begin{figure*}
\centering
\includegraphics[width=\linewidth]{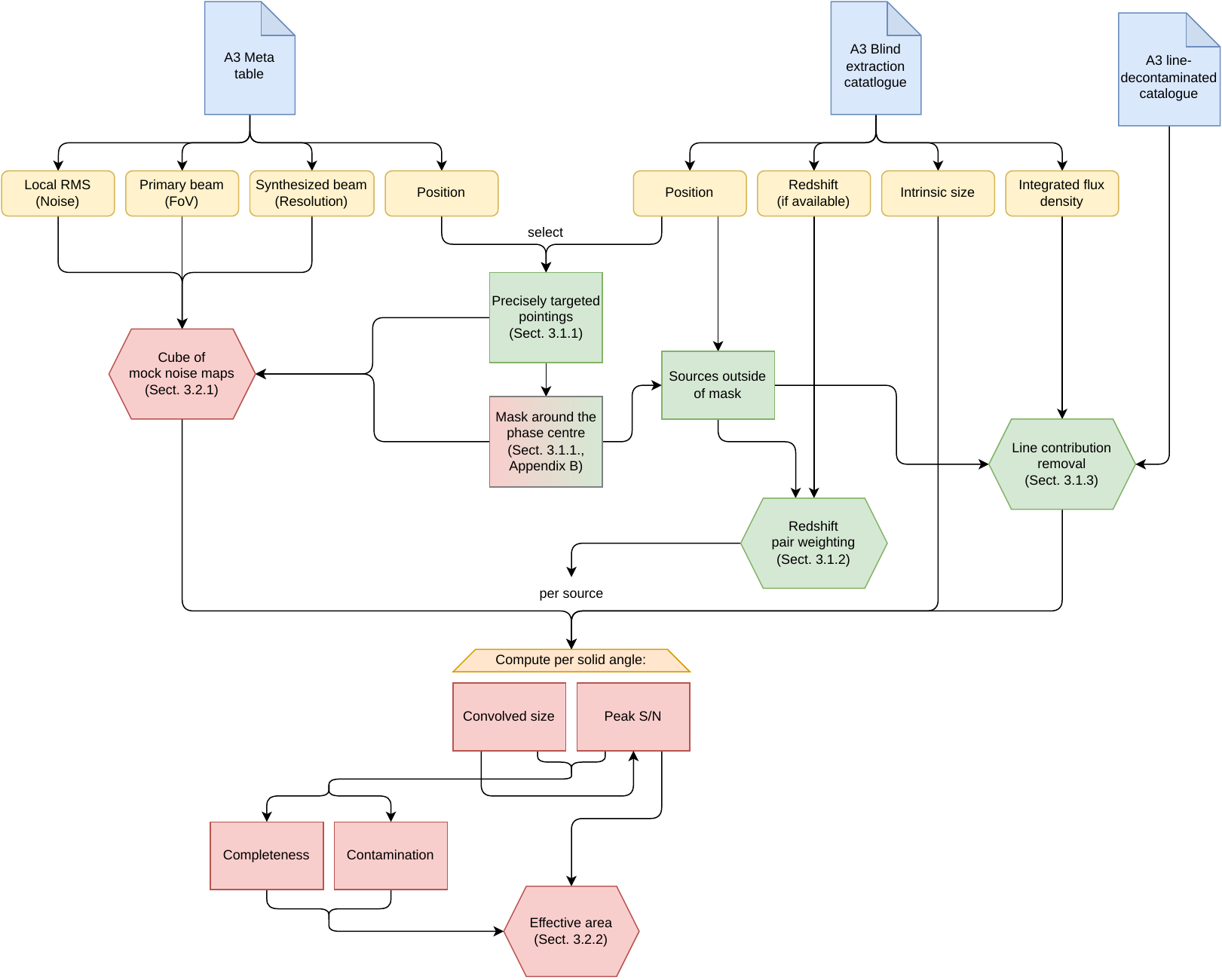}
\caption{Schematic of the process for  retrieving continuum number counts from the inhomogeneous \acosmos database. The red panels are steps in the computation of the effective area, as described in Sect. \ref{subsec_effective_area}; the green panels are steps to correct for selection biases, as described in Sect. \ref{subsec_selection_criteria}. This process is repeated once for each ALMA band.}
\label{flowchart_nc}
\end{figure*}

Galaxy number counts are a useful measurement to constrain galaxy evolution models and the longer wavelength regimes covered by ALMA are especially valuable to constrain the population of high-redshift DFSGs in these models \citep[e.g.][]{casey18b, casey18}. 
Number counts within different ALMA bands have been established in the past \citep[see e.g.][]{karim13,oteo16,stach18,hatsukade18,gonzales-lopez19, gonzales-lopez20, simpson20,bethermin20,zavala21,gomez-guijarro21,chen23}.  
Ideally, number counts are obtained from blind surveys, which are characterised by a large contiguous area and the absence of selection biases. 
Although containing a number of blind surveys, our \acosmos database is an intrinsically inhomogeneous accumulation of observations from a large number of projects with different scientific objectives, tailored to the particular targets and requirements set by each individual PI. 
This circumstance negates both advantages inherent to blind surveys and introduces two major challenges: 
removing selection biases and determining the correct effective sky area. 
The first challenge is associated with the fact that most of the images in the \acosmos database are targeted on specific sources.
Single pointings are targeted observations as a rule and therefore usually have a source at or near their phase centre. 
If these targets are counted, this results in a massive overestimation of the source surface density. 
The second challenge is associated with the inhomogeneous nature of the database.
To determine the effective area of this `survey', the differences in wavelength, depth, and resolution between observations as well as their spatial overlap need to be properly taken into account. 

There have been past attempts to infer number counts from single-pointing-based data with ALMA \citep[][]{oteo16,zavala18,bethermin20,chen23}. 
Those studies, however, either deal with few projects or use calibrator observations, so they can reliably exclude their targets manually.
With our large data volume, it is unfeasible do a manual selection.
Therefore, we have to define an algorithm.

In this section, we describe the procedure by which we attempt to derive number counts from the fundamentally heterogeneous \acosmos database \citep[see also][for a similar method]{traina23}. 
The procedure is schematically outlined in Fig.~\ref{flowchart_nc} and consists of two main parts: 
the selection criteria applied to correct for observational biases and achieve a `blinding' of the data, and the calculation of the effective area.

We chose the blind source catalogue as the basis for our number counts analysis in order to also include sources without a known optical counterpart. 
Sources can be listed in this catalogue multiple times if they are covered by multiple pointings or surveys.
We therefore first reduced the catalogue to only one detection per source and ALMA band. 
In cases of multiple detections for one source in a given band, we only considered the detection with the highest significance (i.e. highest S/N$_{\rm peak}$).

\subsection{`Blinding': Source selection} \label{subsec_selection_criteria}

The removal of selection biases from a database in which the scientific objectives of each observation were selected by individual PIs with all types of galaxies is challenging but vital in attempting to measure the underlying source surface density. 
A form of selection and exclusion has to be made to the detected sources in order to correct for those biases. 
The most important factor is the identification and exclusion of observational targets.
Additionally, sources physically associated with those targets due to clustering can cause an artificial excess in the number counts. 
While clustering is an integral part of the distribution of sources in the sky, intentional observation of clusters \citep[e.g.][]{umehata17} and/or bright sources, which have an increased probability of being in a cluster environment, can result in artificially high source counts \citep[e.g.][]{gruppioni20}. 
In the following sections, the selection criteria to minimise those biases are described and tested via simulations.

\subsubsection{Offset and masking} \label{subsubsec_masking}

Identifying the target of single pointings without prior know\-ledge on the observational goals of the respective projects poses a big challenge. 
Single pointings are naturally aimed at the position of a known source, the target is thus expected to be exactly at the phase centre of the respective pointing. 
However, the known prior position can be offset from the source position when observed with ALMA, for example, if it is a follow-up of submillimetre galaxies selected from single-dish surveys which are characterised by large positional uncertainties \citep[$\gtrsim$3\,arcsec;][]{hodge13}. 
From the \acosmos images alone, observational targets offset from the phase centre are indistinguishable from serendipitous sources that happen to be in the field of view, especially if a source is offset only by a few arcseconds from the phase centre.  
As there are $\sim$230 unique projects and $\sim$4000 images in the entire database, it is impossible to examine individually every image considering the respective project in order to determine which source/s was/were the intended target/s in each individual case. 

Instead of making uncertain assumptions about which sources can or cannot be considered targets, we chose a more conservative approach \citep[see also][]{traina23}: 
we limited our analysis to what we  refer to in the following as `precisely targeted' single pointings that have a source right at their phase centre (i.e. the extracted position is within a radius of 1\,arcsec around the phase centre). 
Since the probability of a source being serendipitously detected within this small area is low (of the order of $\sim$0.1\%), we have a degree of certainty to say that these sources are actual targets and should be excluded from our number counts analysis. 
To exclude these targets from our analysis as well as further potential contaminants close to them due to clustering, we masked a central circular area around the phase centre of each precisely targeted pointing. 
All sources inside the mask were removed from our analysis and the corresponding area was excluded from the calculation of the effective area (Sect.~\ref{subsubsec_noise_map}). 
The radius of this mask was chosen to be the smallest radius beyond which the number counts do not significantly change when the radius is further increased.  
In Bands~3, 4, and 6 the mask radius was set to 1\,arcsec, in Band~7 to 4\,arcsec. 
The choice of the mask was also tested through our simulations described in Sect.~\ref{subsec_sim} (see Appendix~\ref{appendix_mask} for details).

It would be intuitive to assume that masking out all (usually bright) target sources would in turn introduce a negative bias to the bright end of our number counts.
However, the sky distribution of galaxies follows a Poisson distribution (not considering clustering) and thus the probability of detecting a bright source in a given line of sight (LOS) is independent of the presence of another bright source in its vicinity. 
Therefore, bright sources are still detected serendipitously and the bright end of the number counts can still be constrained. 
The only limitation to this is clustering, which we accounted for by considering the redshift information of targets and serendipitous sources (see Sect.~\ref{subsubsec_pairs}).
Our simulations (see Sect.~\ref{subsec_sim}) show that using this approach, even when targeting only the brightest sources in a given field and masking them out, it does not introduce a negative bias on the bright end of the number counts.

In blind surveys, no specific selection or masking is needed since there are no dedicated targets. 
Thus, every source was treated as serendipitous.

\subsubsection{Redshift pairs} \label{subsubsec_pairs}

While a central mask can exclude the most proximate cluster neighbours, the typical angular scale of galaxy clusters (e.g. $\sim$240\,arcsec for a cluster at $z=2$ with a 2\,Mpc diameter) exceeds the primary beam size of typical ALMA pointings (e.g. $\sim$60\,arcsec in Band 3). 
It is therefore necessary to identify among our serendipitous sources (i.e. outside the central mask) those which are physically associated with the target-source (i.e. at the phase centre) by more than purely angular proximity. 

To this end, we made use of the availability of prior redshift information for most detected sources within our \acosmos database. 
The expected maximum redshift difference between two galaxies located within the same cluster is of the order of $\Delta z \approx 10^{-3}$ (for a cluster at $z\lesssim6$ with a 2\,Mpc diameter). 
However, measured redshifts (especially when measured photometrically) are subject to uncertainties that often exceed this order of magnitude. 
It is therefore not possible to make absolute statements about two sources being associated in the same cluster, and instead we have to assess the likeliness of such an association.
To do so, we used the integral of the overlap between the redshift probability distribution function (PDF) of a serendipitous source and its corresponding target-source as an estimate of the likelihood $P_{\textup{pair}}$ of both sources having the same redshift and thus being located within the same galaxy cluster. 
For sources with only a photometric redshift estimates we used as PDFs a Gaussian function centred on their redshifts with a width of  $0.06(1+z)$, that is, the median photometric redshift uncertainty in the COSMOS2020 catalogues \citep[][]{cosmos2020}. 
For sources with a spectroscopic redshift we used a PDF width of $0.001(1+z)$ \citep[e.g. ][]{fernandez01}.
The contribution of each serendipitous source to the number counts was then weighted by its likelihood to have the same redshift as the target-source by multiplying it with a factor $1-P_{\textup{pair},i}$, where $P_{\textup{pair}}$ ranges from 0, for a very unlikely physical association between the serendipitous source and the target of the pointing, to 1 for a very likely association.  

Naturally, we cannot apply this weighting to a fraction of source pairs ($\sim$20\%) for which no redshift information is available (for either the target or the serendipitous source). 
In that case, $P_{\textup{pair}}$ was simply set to a value of 0, with the exception of one source known to be in a cluster with the observational target, where we set $P_{\textup{pair}}=1$ \citep[CRLE,][]{pavesi18}.

\subsubsection{Line contamination}

There is a small probability that the continuum flux densities of some of sources selected for our number counts may be contaminated by emission from bright lines that happen to fall within the frequency window of the observation. 
To account for this, we made use of the fact that our \acosmos pipeline already includes a routine that calculates the contribution of bright emission lines to the continuum flux densities (see Sect.~\ref{subsec_a3cosmos_pipeline}). 
However, as this requires an SED fit it can only be performed for sources that have prior counterparts and redshift information.
This is the case for almost all serendipitous sources in Band~3 and 7, and for $\sim$60\% of the serendipitous sources in Band~4 and 6. 
We find a significant contribution of line emission to the continuum (i.e. $\geq10\%$ of the measured integrated flux density) for one source in Band~3 and for $\sim$25\% and $\sim$10\% of the sources in Band~4 and 6, respectively. 
No significant line contribution is found in Band~7. 
Despite this rather low degree of line contamination, we decided to use the line-decontaminated total flux densities to calculate the effective area associated with each galaxy (see Sect.~\ref{subsec_effective_area}). 
Additionally, we excluded sources from our number counts analysis if their line-decontaminated peak flux density fell below our minimum threshold of S/N$_{\rm peak}=5.4$.
This applied only to one source in Band~3, two sources in Band~4, and two sources in Band~6.
We note that although we chose to correct for line contamination for the sake of thoroughness, this does not significantly affect the results of our analysis as the differences are well within the uncertainties.

\subsection{Effective area} \label{subsec_effective_area}

\begin{figure*}
\centering
\includegraphics[width=0.6\linewidth]{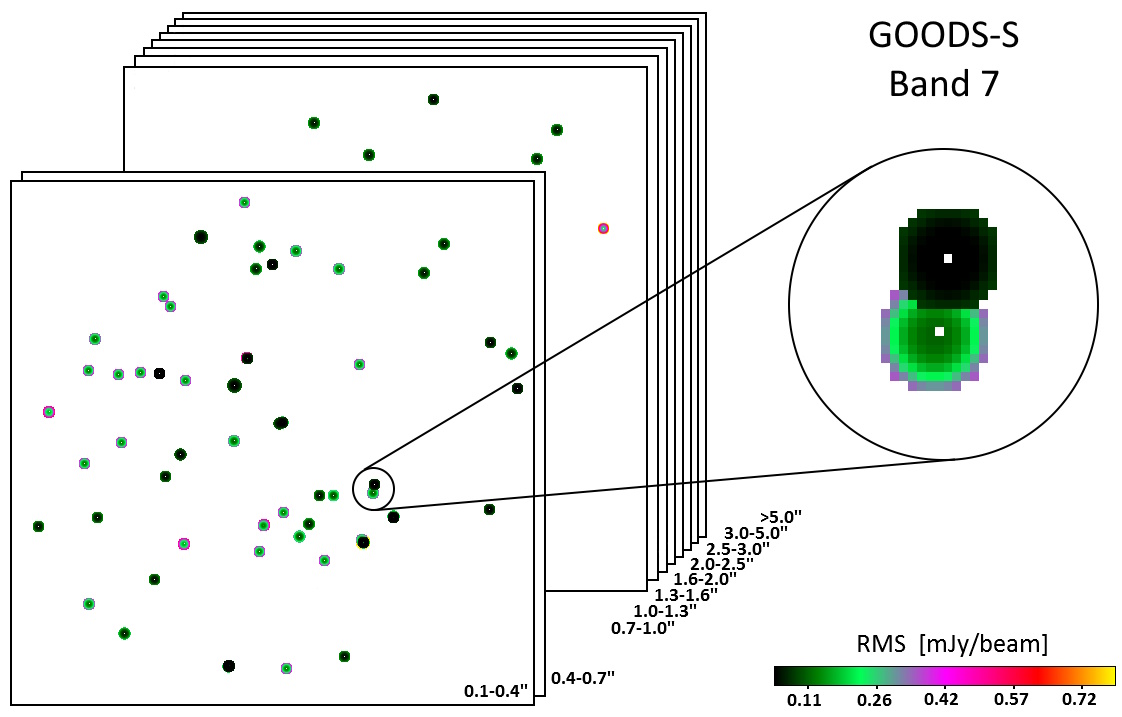}
\caption{Illustration of the noise cube for the GOODS-S field in ALMA Band 7 (see Sect. \ref{subsubsec_noise_map}). Each cube consists of ten layers for different angular resolutions, as indicated in the bottom right of each layer. Each pixel represents a solid angle of $2\times2$\,arcsec. The colour of each pixel indicates the lowest RMS noise available between all pointings covering this solid angle in that particular band and at this angular resolution. The enlarged cutout shows how, as a result, pointings with lower RMS overwrite those with higher RMS in cases of spatial overlap between pointings with similar angular resolution.}
\label{cube_example}
\end{figure*}

Number counts can be calculated by inversely summing the effective area of all sources above a given flux density $S$:

\begin{equation}
    N(>S) = \sum_i \frac{1}{A_{\textup{eff}}(S_i, \theta_i)} \label{eq_numbercounts} ,
\end{equation}

\noindent
When introducing pair-weighting (see Sect. \ref{subsubsec_pairs}), this equation becomes: 

\begin{equation}
    N(>S) = \sum_i \frac{1 - P_{\textup{pair},i}}{A_{\textup{eff}}(S_i, \theta_i)}. \label{eq_numbercounts_weighted}
\end{equation}

\noindent
Here, the effective area of a source, $ A_{\textup{eff}}$, is given by 

\begin{equation}
 A_{\textup{eff}}(S_i, \theta_i) = \iint_{\textup{Survey}}  \frac{C_{\textup{compl.}}(S_i, \theta_i, x, y)   } { 1 - C_{\textup{contam.}}(S_i, \theta_i, x, y)  } D(S_i, \theta_i, x, y)   \, d\Omega  \label{eq_aeff_int} , 
\end{equation}

\noindent
and depends on the detectability $D$, the completeness $C_{\textup{compl.}}$ and the contamination $C_{\textup{contam.}}$\footnote{The contamination is an equivalent measure to the commonly used `purity', which corresponds to $1-C_{\textup{contam.}}$. Here, we use contamination, as it is directly obtained from the spurious fraction (see Sect. \ref{subsubsec_read_map}).}. 
A source is considered detectable if its S/N$_{\rm peak}$, which depends on the total source flux density $S_i$ and the beam-convolved source size $\theta_i$, at this particular position is greater than the blind extraction threshold of 5.4 (see Sect. \ref{subsec_a3cosmos_pipeline}), therefore $D = 1$ where S/N$_{\textup{peak}}>5.4$ and $D=0$ otherwise.   
The completeness (i.e. the recovery rate of real sources) and contamination (i.e. the like\-lihood of a source being a spurious detection) both depend on S/N$_{\rm peak}$ and $\theta_i$ and range from 0 to 1. 
For any given source, the detectability, completeness and contamination thus vary largely between different observations, depending on wavelength, sensitivity and resolution of the respective images. 
A two-step approach was thus chosen in order to incorporate the effect of detectability, completeness and contamination on the calculation of the effective area of a given source: 
we first created noise coverage maps of the COSMOS and GOODS-S fields for each ALMA band into which we wrote the positions and sensitivities of each ALMA image, split into bins of different spatial resolution and prioritising low over higher RMS noise in areas of spatial overlap. 
We then used these maps to determine the detectability, completeness and contamination for each source in each solid angle of the COSMOS and GOODS-S fields and thereby calculate its associated effective area. 

\subsubsection{Noise maps} \label{subsubsec_noise_map}

In order to assess the detectability of a source of a given flux density and size in each solid angle of the COSMOS and GOODSS fields, we need to take into account both the RMS and spatial resolution of all images available in our database. 
While the detectability of point-like sources depends only on the image sensitivity, an extended source is more easily detectable in a low angular resolution image than in a high angular resolution image with the same sensitivity per beam.  

To account for the effect of both RMS and angular resolution, we chose the following approach:
For each ALMA band, we set up empty maps of the COSMOS and GOODS-S fields with a pixel-scale of 2\,arcsec. 
Each map has ten layers corresponding to different angular resolutions, which are: 
0.1-0.4'', 0.4-0.7'', 0.7-1.0'', 1.0-1.3'', 1.3-1.6'', 1.6-2.0'', 2.0-2.5'', 2.5-3.0'', 3.0-5.0'' and >5''. 
These bins were chosen so that the spatial overlap of ALMA pointings within each layer is minimised. 
The RMS of all available images from the database ($\sigma_{\nu}$) was normalised to the central frequency of its respective band ($\sigma_{\rm \nu, ref}$) under the assumption of being in the regime of the Rayleigh-Jeans tail of the dust continuum emission and choosing a dust spectral index $\beta=1.8$ \citep[e.g.][]{scoville14}:

\begin{equation}
\sigma_{\nu,\, \textup{ref}} = \sigma_{\nu} \cdot \left( \frac{\nu_{\textup{ref}}}{\nu} \right)^{3.8} .
\end{equation}
These normalised RMS maps were then corrected for the primary beam attenuation, causing the noise to increase with distance from the phase centre. 
Areas beyond a  primary beam response of 0.3 were disregarded as well as a central radius of 1 or 4\,arcsec around the phase centre of single pointings (Sect. \ref{subsubsec_read_map},  \ref{subsec_selection_criteria} and Appendix \ref{appendix_mask}).
The renormalised primary-beam corrected RMS maps were then written into the layer corresponding to their angular resolution of the respective field maps at the positions corresponding to their sky coordinates. 
In case of spatial overlap between pointings of similar angular resolution, lower RMS was favoured over higher RMS.
In addition to this so-created `noise cube', the resolution properties of all images (that is, major and minor axis as well as position angle of the synthesised beam) were stored in equivalent map cubes.
As a result, we have for each ALMA band four ten-layered map cubes per field, one storing the normalised noise and three storing the corresponding resolution information. 
These cubes are the basis on which the effective area for each source was calculated.
An example of such a noise cube for Band 7 in the GOODS-S field is shown in Fig. \ref{cube_example}.
This example illustrates the primary-beam corrected noise profiles, the masking, and the treatment of spatial overlap between two pointings of similar angular resolution in our calculation of the effective area.

\subsubsection{Computing the effective area} \label{subsubsec_read_map}

The sources chosen for our number counts analysis are blindly detected sources with S/N$_{\textup{peak}}\ge 5.4$. 
Since the edges of images are more likely to be contaminated with spurious sources due to a higher and not well-behaved noise, we only considered the area of images where the primary beam attenuation is >0.3 and used only sources found therein.
Blind source extraction provides the integrated flux density $S_{\textup{total}}$ and peak flux density $S_{\textup{peak}}$ of each source.  
From this, the circularised intrinsic source size, $\theta_{\rm intr.}$, was backwards-inferred by dividing $S_{\textup{total}}$ by $S_{\textup{peak}}$ and performing a deconvolution with the circularised size of the synthesised beam of the image in which the source was originally detected, $\theta_{\rm beam}$:

\begin{equation}
\theta_{\rm intr.} = \theta_{\rm beam}  \cdot \sqrt{ \frac{S_{\rm total}}{S_{\rm peak}} -1}.
\end{equation}

\noindent
We chose to backwards-infer the intrinsic source size from the flux density measurements, rather than from the \texttt{PyBDSF} 2D Gaussian modelling, as it allowed us to consistently predict the measured S/N$_{\rm peak}$ of each source in their respective original image.
While in most cases this is also works with the source size information provided by \texttt{PyBDSF}, there are also rare cases in which it is not possible, for example when a source is fitted by \texttt{PyBDSF} with two or more Gaussian profiles.
Sources extracted from low angular resolution images are typically retrieved as unresolved (i.e. $\theta_{\rm intr.}=0$). 
To account for the fact that such sources could be resolved in images of higher angular re\-so\-lu\-tion, we assigned them a typical intrinsic source size of radius $r=0.1\pm0.05$\,arcsec, following the findings of \citet{gomez-guijarro21}.

We used $\theta_{\textup{intr.}}$ and $S_{\textup{total}}$ to compute the expected detectability of each source in a given LOS of the COSMOS and GOODS-S fields, using

\begin{equation}
\textup{S/N}_{\textup{peak,LOS}} = \frac{S_{\textup{total},\nu} \cdot \left( \frac{\nu_{\textup{ref}}}{\nu} \right)^{3.8}}{\frac{\theta_{\textup{intr.}}^2+\theta_{\textup{beam,LOS}}^2}{\theta_{\textup{beam,LOS}}^2} \cdot \textup{RMS}_{\textup{LOS},\nu_{\textup{ref}}}},
\end{equation}

\noindent where $\nu$ is the observed frequency, $\nu_{\textup{ref}}$ is the central frequency of the band, $\theta_{\textup{\rm beam,LOS}}$ is the circularised beam axis of the most sensitive observation in that angular resolution layer and $\textup{RMS}_{\textup{LOS}}$ is the corresponding RMS value. 
If the expected peak signal-to-noise was above the minimum threshold of 5.4, the source was considered detectable at that angular resolution and within the solid angle intersecting the area $A_{\textup{LOS}}=4\,\textup{arcsec}^2$ (i.e. the previously defined pixel-scale of 2\,arcsec of our noise cubes).

Using S/N$_{\textup{peak,LOS}}$ and $\theta_{\textup{conv,LOS}}=\sqrt{\theta_{\textup{intr.}}^2+\theta_{\textup{beam, LOS}}^2}$ the correction factors for completeness $C_{\textup{compl,LOS}}$ and contamination $C_{\textup{contam,LOS}}$ were determined. 
The completeness correction for a given S/N$_{\textup{peak,LOS}}$ and $\theta_{\textup{conv,LOS}}$ was adopted from \citet{a3cosmos_1} who inferred these corrections using Monte Carlo simulations in which artificial sources were inserted into residual images and recovered with the same source extraction method that was used to build the \acosmos blind catalogue. 
As a rule, the completeness increases with S/N$_{\textup{peak}}$ and is slightly higher for less extended sources.
Above our detection threshold of S/N$_{\textup{peak}}>5.4$, the correction is small as the completeness is always $\gtrsim70\%$. 
The contamination correction was determined from the fraction of spurious sources.
This spurious fraction was obtained by dividing the number of detections in the \acosmos negative ALMA images (i.e. original \acosmos images multiplied by $-1$, so any detection is by definition spurious) by the number of detections in the positive (i.e. original) \acosmos images (i.e. consisting of both real and spurious sources). 
This spurious fraction decreases with both signal-to-noise and source size. 
Again, for most cases the correction is small, that is, on average $\sim$20\% at S/N$_{\textup{peak}}=5.4-6.0$ and <15\% for S/N$_{\textup{peak}}>6.0$.
The corrections for completeness and contamination were directly applied to the area associated with a given solid angle:

\begin{equation}
    A_{\textup{eff,LOS}} = \frac{A_{\textup{LOS}} \cdot C_{\textup{compl,LOS}} }{ 1 - C_{\textup{contam,LOS}}}.\label{eq_aeff_pix} 
\end{equation}
With this approach, the effective area for a given source in a given angular resolution layer is the sum of $A_{\textup{eff,LOS}}$ over all lines of sight in the COSMOS and GOODS-S noise cubes. 
Finally, for sources detectable in a given solid angle in more than one angular resolution layer, we counted the layer with the highest value of $A_{\textup{eff,LOS}}$. 
Naturally, this calculation of the effective area was made independently for each band (i.e. to infer the number counts in a given band, we only account for the source detected in this band and considering the noise cubes of this band).

\subsection{Simulations} \label{subsec_sim}

In order to test the applicability of our selection criteria, and to some degree also our effective area calculation, to accurately retrieve number counts from a biased database such as ours, we used simulations to recreate a biased sample of observations and infer number counts with our previously described method.

As basis for these simulations, we used the Simulated Infrared Dusty Extragalactic Sky (SIDES) from \citet[][see also \citealt{bethermin17}]{gkogkou23}. 
SIDES offers 117 unique patches of simulated sky of 1\,deg$^2$ each, accounting for clustering and including information on redshift and continuum flux density in all relevant ALMA bands. 
Since the COSMOS and GOODS-S fields together cover an approximate area of 2\,deg$^2$, we took 116 of those patches and combined them into independent 58 realisations of a 2\,deg$^2$ field. 
 
\begin{figure}[ht!]
\centering
\includegraphics[width=\linewidth]{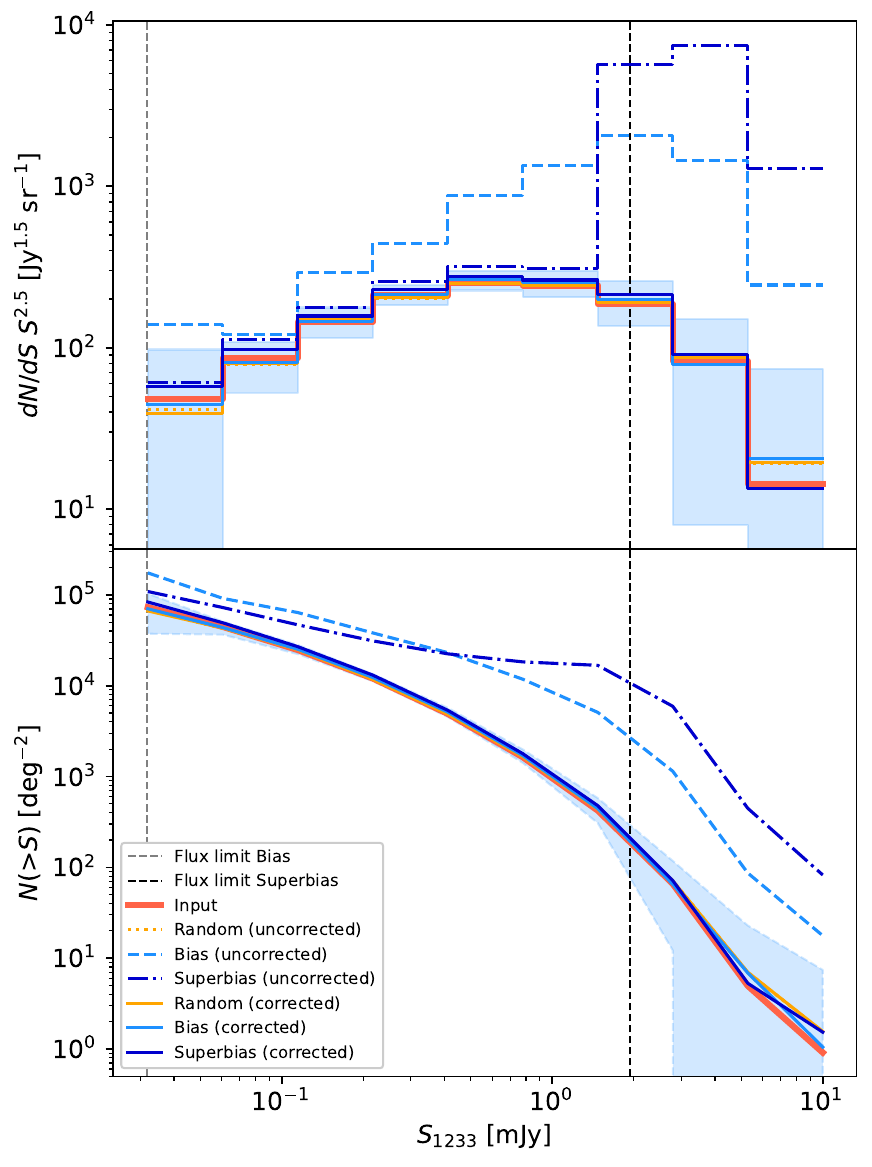}
  \caption{Simulated differential (\textit{top}) and cumulative (\textit{bottom}) number counts for single pointings in ALMA Band 6 based on the SIDES simulated sky catalogue \citep[][]{bethermin17,gkogkou23}, the pointing distribution of \acosmos, and our applied selection criteria. The coloured lines show the number counts for different simulated biases (see Sect. \ref{subsec_sim}). Dark orange indicates the input number counts as inferred from the complete SIDES catalogue. Light orange is for the pointings randomly distributed on the sky (Random). Dark blue is for the pointings that target all the brightest sources (Superbias). Light blue is for the pointings that  reflect the target source brightness distribution in \acosmos (Bias). The dashed, dash-dotted, and dotted lines are number counts inferred from the complete pointings; the  solid lines show the number counts when applying a central mask of 1\,arcsec and weighting by redshift proximity. The shaded region shows the $1\,\sigma$ dispersion between all realisations of the corrected Bias line, but is approximately equal for all corrected lines.
  The vertical black and grey dashed lines indicate the flux density of the faintest target source in the Superbias and Bias cases, respectively. 
  We note that the uncorrected Random line is not visible in the lower panel, as it fully agrees with the corrected Random line.}
\label{sim_b6}
\end{figure}

In each of these 58 fields (i.e. per realisation), we made mock observations by placing in them mock ALMA pointings imitating the set of precisely targeted pointings used in our real number counts analysis.
We then retrieved number counts from these mock observations using the SIDES catalogue to compute the S/N$_{\rm peak}$ of the mock sources located inside these mock pointings.  
For a detailed description of these mock observations and the number counts retrieval see Appendix \ref{appendix_sim}.
We used three modes of simulations differing in the placement of their mock pointings (i.e. choice of mock target sources) to simulate different observational biases: `Superbias', `Bias', and `Random'.
The Superbias mode models the extreme case in which only the brightest sources in a given field have been targeted by the ALMA observers.
The Bias mode represents a more realistic case as the mock targets are chosen based on the flux density distribution of the real observational targets in our pointings. 
Finally, the Random mode serves as a sanity check, as the number counts are measured on a set of mock pointings not targeting any mock sources, distributed randomly instead.

In all three modes, we used two approaches to retrieve the mock number counts: 
the `uncorrected' number counts are retrieved accounting for the complete areal coverage of the mock pointings and all sources within;
the `corrected' number counts are inferred after applying the same corrections as in the real case (i.e. masking and weighting).
Per realisation, the number counts were computed separately for all bias modes, in both the uncorrected and corrected cases.
The final recovered mock number counts are the mean over all 58 realisations. 
Variation from one realisation to another is characterised through the dispersion over these 58 realisations and corresponds to the cosmic variance affecting our estimates. 
We verified that this cosmic variance is dominated by pure Poisson error at the areal coverage of our database.

The Band 6 mock number counts inferred using these simulations and with a central masking radius of 1\,arcsec radius are shown in Fig.~\ref{sim_b6}. 
These are compared to the `Input’ number counts which correspond to the number counts inferred using the complete input SIDES catalogue. 
Because the cosmic variance is similar for all the different modes and number counts retrieval approaches, we show for clarity the dispersion only for the corrected Bias line. 
The Random uncorrected and corrected number counts are in good agreement with the Input number counts. 
This is expected, as in the absence of observational bias this set of mock pointings behaves effectively as a blind survey, only affected by cosmic variance.
Applying additional masking has no further impact in that case other than slightly decreasing their effective area and thus increasing the cosmic variance. 
The Random mode validates our measurement of the effective area.
The uncorrected Bias and Superbias number counts show an expected large excess over the real distribution, up to two orders of magnitude for the bright end in Superbias mode and still more than one order of magnitude in Bias mode. 
The excess in the uncorrected Superbias mode is predominantly at the bright end, since all existing sources above a certain flux density limit are targeted and counted. 
This limit is indicated by a vertical line in Fig.~\ref{sim_b6}. 
Below this limit, however, there is still a notable, albeit much smaller, systematic excess (by $\sim$25\%), which is due to clustering, as massive and bright galaxies at $z>1$ are more likely to be in an over-dense environment (in SIDES and reality) and therefore have a higher number of less bright neighbours. 
The excess of the uncorrected Bias number counts is spread over the entire flux density range, albeit largest at the bright end. 
This bias probably reflects the actual situation in the \acosmos database, in which bright sources are often favoured as observational targets, but fainter sources are also being investigated. 
The effect of clustering here is not as obvious, because the target bias dominates over the whole flux density range. 
These uncorrected number counts demonstrate that it is indispensable to apply a form of masking in attempting to extract the underlying real number counts from these biased observations. 

In the corrected Bias and Superbias modes, there is a clear improvement compared to the respective uncorrected number counts. 
The remaining systematics affect the number counts at a level which is an order of magnitude lower than the uncertainties associated with the cosmic variance (see Appendix~\ref{appendix_mask}).
We notice that these uncertainties increase towards the bright end due to the lower number of sources present in this flux density range, and towards the faint end due to the decrease in effective areal coverage. 

The good agreement of the corrected Superbias number counts with the Input number counts also shows that, as previously mentioned, masking out targets does not prevent constraining the bright end of the number counts, even if all brightest sources are targeted.
Since sources are Poisson distributed, bright sources, while masked out as targets, are also occasionally detected as a serendipitous sources in other pointings and thus still enter our number counts analysis.
 
We conclude from these simulations that with a central mask and redshift pair weighting applied, an ensemble of single pointings is able to yield accurate (with a mean difference of no more than $\sim$10\%) number counts.

\section{Results on the number counts} \label{sec_results}

\begin{table*}[ht!]
  \caption[]{Number of images available to infer the number counts before and after applying our selection criteria. }
     \label{table-selection}
 $$ 
     \begin{tabular}{lcccc}
        \hline
        \noalign{\smallskip}
        ALMA Band   & Blind survey images & Single pointings & \makecell[c]{Precisely targeted \\ pointings}  & \makecell[c]{With serendipitous \\ detection }\\
        \noalign{\smallskip}
        \hline
        \noalign{\smallskip}
        Band 3 &  5  &  413 &  68 &    7   \\
        Band 4 &  2  &  158 &  75 &   16   \\
        Band 5 &  0  &   12 &   0 &    0   \\
        Band 6 & 24  & 1319 & 361 &   60   \\
        Band 7 &  0  & 1926 & 734 &   94   \\
        Band 8 &  0  &   42 &  12 &    0   \\
        Band 9 &  0  &   19 &   6 &    3   \\             
        \noalign{\smallskip}
        \hline
     \end{tabular}
 $$ 
Notes: For each ALMA band the total number of single pointings and blind survey images is given (the latter not affected by selection bias). The next columns list the number of single pointings that contain a detected source at a distance $<1$\,arcsec from the phase centre (called the  precisely targeted pointings) and among those the number of pointings that contain at least one serendipitous detection outside of the central mask of 1\,arcsec radius (Band 7: 4\,arcsec).

\end{table*}

Having demonstrated that our selection bias correction method can successfully recover the number counts in simulations (see Sect.~\ref{subsec_sim}), we apply the same technique to our real data and infer the number counts.
Table~\ref{table-selection} gives an overview of the total number of single pointings available per band in our database compared to the number that remains after excluding pointings without a detection at their phase centre. 
The table also lists the number of precisely targeted pointings that have a serendipitous detection outside of their central mask. 
Figure~\ref{sensitivity_curve} shows as dashed lines the areal coverage of these precisely targeted pointings and mosaics, after cropping the edges and masking out the phase centres. 
The total area available in Band~3 is reduced by $\sim$75\%, in Band~7 by $\sim$70\% and in Band~6 by $\sim$63\%. 
Band 4~loses only $\sim$24\% of total available area due to the fact that one large, albeit rather shallow, blind survey dominates the areal coverage in this band.
Unfortunately, in Bands~5, 8, and 9, we do not have sufficient statistics in terms of number of precisely targeted pointings and serendipitous detections to obtain meaningful constraints on the number counts. 
We do not consider these bands in the rest of our analysis.

\subsection{Single pointing and blind mosaic number counts}

\begin{figure*}
\centering
\includegraphics[width=0.8\linewidth]{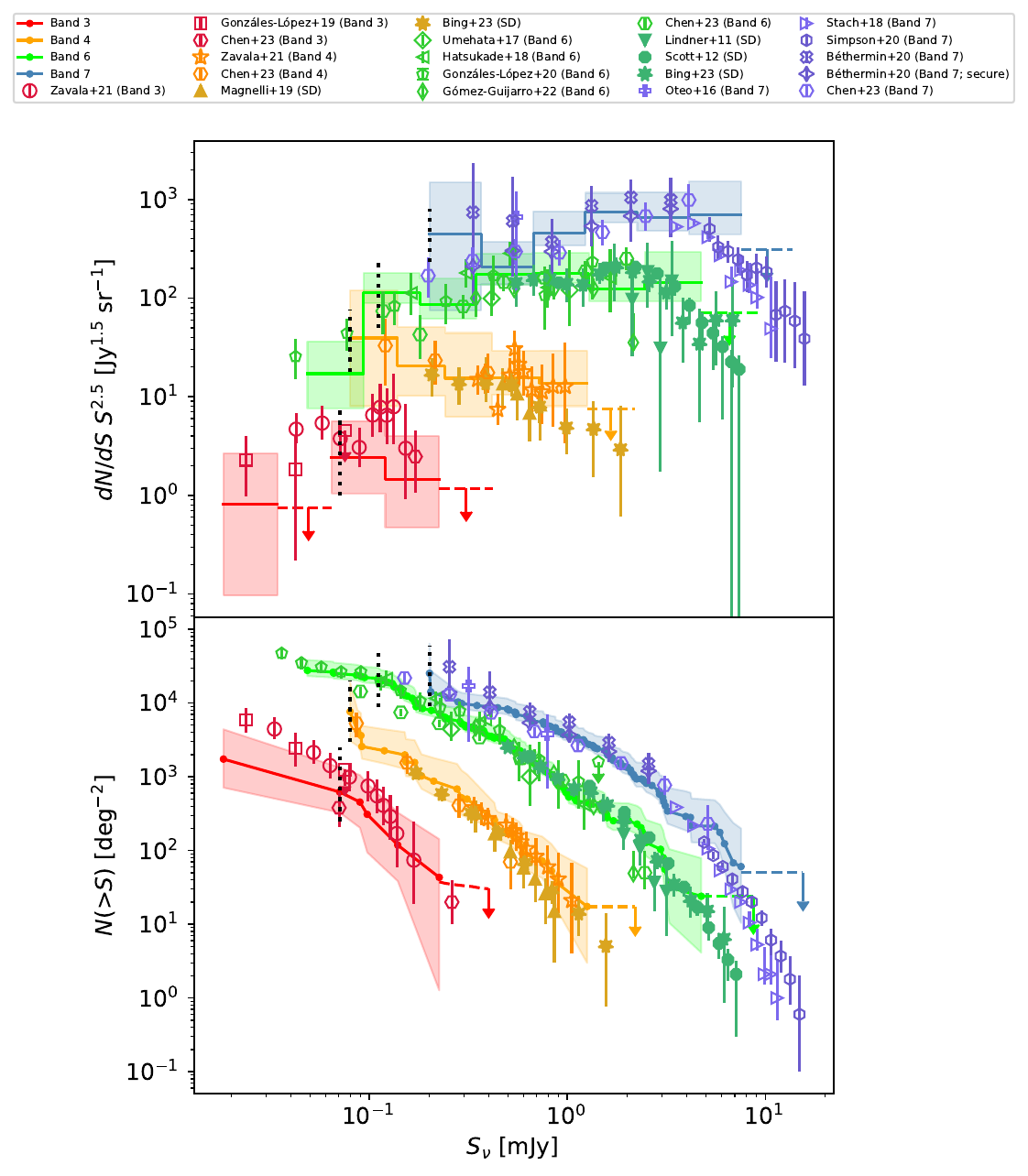}
  \caption{Number counts inferred from the \acosmos database in four ALMA bands: red, Band 3 (2.6--3.6\,mm); orange, Band 4 (1.8--2.4\,mm); green, Band 6 (1.1--1.4\,mm); blue, Band 7 (0.8--1.1\,mm). 
  Number counts were measured by combining blind surveys and single pointings containing a detected source within 1\,arcsec radius around the phase centre and excluding a central area of 1\,arcsec radius in Band 3, 4, and 6, and 4\,arcsec in Band 7. \textit{Top panel}: Differential. \textit{Bottom panel}: Cumulative. The number counts calculated in this work are shown as solid lines, with the shaded area indicating the uncertainty resulting from flux density, contamination, and completeness errors, quadratically combined with Poisson errors.  
  The respective flux densities of the individual sources are shown as points in the cumulative panel. 
  The vertical dotted lines indicate the flux density above which our number counts can be considered  complete. 
  Number counts from the literature are shown as symbols and are colour-coded to the respective ALMA band.}
     \label{figure-nc-all}
\end{figure*}

\begin{table*}
  \caption[]{Differential number counts inferred from the \acosmos database in ALMA Bands 3 ($2998\mu$m), 4 ($2082\mu$m), 6 ($1233\mu$m), and 7 ($925\mu$m).}
     \label{table-diffcount}
 $$ 
     \begin{tabular}{cc|cc|cc|cc}
        \hline
        \noalign{\smallskip}
        \makecell[c]{$S_{2998}$ range \\ {[}mJy{]}}  & \makecell[c]{$dN/dS\ S^{2.5}$ \\ {[}Jy$^{1.5}$ sr$^{-1}${]} }  & \makecell[c]{$S_{2082}$ range \\ {[}mJy{]}} & \makecell[c]{$dN/dS\ S^{2.5}$ \\ {[}Jy$^{1.5}$ sr$^{-1}${]} } & \makecell[c]{$S_{1233}$ range \\ {[}mJy{]}} & \makecell[c]{$dN/dS\ S^{2.5}$ \\ {[}Jy$^{1.5}$ sr$^{-1}${]} } & \makecell[c]{$S_{925}$ range \\ {[}mJy{]}} & \makecell[c]{$dN/dS\ S^{2.5}$ \\  {[}Jy$^{1.5}$ sr$^{-1}${]} } \\ 
        \noalign{\smallskip}
        \hline
        \noalign{\smallskip}
0.018 - 0.034        & $0.81\substack{+1.87 \\ -0.71}$  & 0.08 - 0.14   &  $ 39 \substack{+80 \\ -31}$  &  0.05 - 0.09   & $ 17\substack{+19  \\  -9}$   & 0.20 - 0.37   & $449\substack{+1040 \\ -374}$ \\
0.034 - 0.064        & $<0.75$                                   & 0.14 - 0.24   &  $ 21 \substack{+30 \\ -10}$           &  0.09 - 0.18   & $ 114\substack{+66  \\ -25}$       & 0.37 - 0.67   & $206\substack{+168  \\  -68}$ \\
0.064 - 0.120        & $2.41\substack{+3.28 \\ -1.36}$           & 0.24 - 0.42   &  $ 15 \substack{+29 \\ -9}$               &  0.18 - 0.35   & $ 87\substack{+70  \\ -25}$       & 0.67 - 1.23   & $458\substack{+299  \\ -117}$ \\
0.120 - 0.225        & $1.45\substack{+2.57 \\ -0.98}$           & 0.42 - 0.72   &  $ 16 \substack{+14 \\ -5}$   &  0.35 - 0.67   & $174\substack{+108 \\ -38}$                & 1.23 - 2.25   & $739\substack{+428  \\ -155}$ \\
0.225 - 0.422        & $<1.18$                                   & 0.72 - 1.26   &  $ 14 \substack{+16 \\ -6}$               &  0.67 - 1.28   & $178\substack{+110 \\ -35}$        & 2.25 - 4.12   & $658\substack{+532  \\ -178}$ \\
                    &                                           & 1.26 - 2.18   &  $<8$                                     &  1.28 - 2.47   & $123\substack{+89  \\ -30}$               & 4.12 - 7.54   & $700\substack{+812  \\ -256}$ \\
                    &                                           &                &                                        &  2.47 - 4.75   & $144\substack{+149 \\ -51}$                & 7.54 - 13.8   & $<310$ \\ 
                    &                                           &                &                                        &  4.75 - 9.13   & $<72$                                         &               &  \\  

        \noalign{\smallskip}
        \hline
     \end{tabular}
 $$ 

\end{table*}

Using the method validated by our simulations, we inferred the number counts by combining precisely targeted pointings and blind surveys. 
The results are listed in Table~\ref{table-diffcount} and shown in Fig.~\ref{figure-nc-all} as solid lines.
The number counts are shown in both their differential ($dN/dS$) and cumulative ($N(>S)$) forms, together with number counts from the literature colour-coded by the wavelength of the corresponding ALMA band. 
Errors, shown as shaded regions, are the quadratic combination of the Poisson errors \citep[using approximations from][]{gehrels86} with errors due to flux density uncertainties and errors due to uncertainties on the completeness and contamination corrections.
Since our simulations indicate that the cosmic variance is dominated by pure Poisson error (see Sect.~\ref{subsec_sim}), no additional cosmic variance uncertainty was added to our error bars.
The cumulative counts are shown at the measured total flux densities of each source in a band.   
The dashed horizontal lines indicate upper limits for the number counts above the flux density of the brightest source in a given band. 

Significantly extended sources with low total flux densities are difficult to detect due to a low S/N$_{\rm peak}$ in images with a high or moderate angular resolution. 
Therefore, our database holds the risk of insufficiently detecting such sources, introducing an additional incompleteness effect, which lowers the accuracy of our number counts in lower flux density ranges. 
To see if this issue considerably affects our results, we defined the size of a significantly extended source (a circularised full width at half maximum of $\sim$0.66\,arcsec, which corresponds to the 80th percentile of the distribution of circularised source radii in our blind catalogue) and determined the lowest total flux density at which a source of such size is detected in our source sample for each band. 
This limit is marked by the black vertical dotted lines in Fig. \ref{figure-nc-all}.
Band~3 and 6 are potentially affected in their low flux density bins $\lesssim0.1$\,mJy.

\subsection{Comparison with previous ALMA number counts}

In Band 3, we first compare our results to the ASPECS-LP ALMA 3\,mm number counts \citep{gonzales-lopez19} and find that they are by a factor $\sim$3 higher than ours. 
As the combined map from this survey is contained in our database (see Sect. \ref{subsec_comb_mosaics}), this disagreement is somehow surprising. 
Therefore, we directly compared the sources extracted from this map with the six continuum detections listed in \citet{gonzales-lopez19}.
Only their brighter source (`CO1') was recovered above our detection threshold of S/N$_{\rm peak}=5.4$ and at a significantly lower flux density than listed in their work (18 vs  33\,$\mu$Jy). 
This source, as well as the other five, are also identified in \citet{gonzales-lopez19} as CO line emitters.
This lead us to suspect that the continuum fluxes listed in and used in the number counts of \citet{gonzales-lopez19} were not measured on the publicly available `linefree' continuum map (see Footnote \ref{footnote_aspecs}) but from a different version still containing line emission.
To test this, we ran our prior-based \texttt{GALFIT} source extraction on the five frequency coverages of this survey, as downloaded from the ALMA archive and processed by our pipeline (see Sect. \ref{subsec_a3cosmos_pipeline}).
Three out of the six continuum sources listed in \citet{gonzales-lopez19} were recovered at S/N$_{\rm peak}>3$: `CO1', `CO2', and `CO3'. 
The source `CO1' was found in four of these frequency coverages.
In two of them, the \texttt{GALFIT} flux densities agree with our measurement on the `linefree' map, while in the other the flux density is significantly higher (i.e. 56 and 87\,$\mu$m). 
The frequencies covered by these two latter tunings correspond to that of the CO line for this galaxy listed in \citet{gonzales-lopez19} (CO\,3-2 at $z=2.543$).
The other two sources were recovered in only one frequency coverage, also corresponding to their respective CO lines (CO\,3-2 at $z=2.696$ and CO\,2-1 at $z=1.550$). 
These findings support our initial suspicion that the continuum flux densities of the sources listed in \citet{gonzales-lopez19} are based on a map that still contains line contribution. 
Therefore, our number counts can be considered as more accurate than those of \citet{gonzales-lopez19}.

We then compare our Band 3 number counts with the results from \citet{zavala21} who, like us, used ALMA archival data (including the 3\,mm ASPECS-LP) but where the targets and physically associated sources were masked out manually.
At flux densities $\gtrsim0.06$\,mJy, our number counts are consistent within the error bars with their results.
However, at the lower end, their number counts are higher by a factor $\sim$3.
This can be explained by the inclusion of the 3\,mm ASPECS-LP survey in the sample of \citet{zavala21}. 
Like in our analysis, they recovered only one source from the ASPECS-LP map and noted the possible contamination by CO line emission of the flux density of this source. 
For this reason, they replaced the flux density of this source with the one given by \citet{gonzales-lopez19}.
However, as we previously argued, this flux density value is likely still contaminated by line emission.
Fixing the continuum flux density of this source to our measured value would bring the number counts of \citet{zavala21} into agreement, within the uncertainties, with our number counts at the low end.

Lastly, we compare our Band 3 number counts with the results from \citet{chen23}, who used ALMA calibrator observations with masked-out targets. The results are consistent within the uncertainties.

In Band 4, we compare our number counts to the ALMA-based number counts from \citet{zavala21} and \citet{chen23}, and the single-dish counts from \citet{magnelli19} and \citet{bing23}, and find that they agree within the uncertainties. 
We reach a depth of $\sim$0.08\,mJy, similar to the single-pointing-based number counts from \citet{chen23}, which is 0.4\,dex and 0.6\,dex lower than the flux density limits of \citet{bing23} and \citet{zavala21}.
Utilising high-sensitivity single pointings allows for the exploration of much fainter sources than those probed by the relatively shallow blind 2\,mm surveys used in \citet{bing23} and \citet{zavala21}. 
We note that, despite still agreeing within the uncertainties, the number counts inferred here are systematically higher than the single-dish number counts from \citet{magnelli19} and \citet{bing23}. 
This suggests that the resolution limitations inherent to single-dish instruments are not negligible at long wavelengths, even with the application of certain corrections.

In Band 6, we compare our results to number counts from the ALMA blind surveys from \citet[][ASAGAO survey]{hatsukade18}, \citet[][ASPECS-LP survey at 1.2\,mm]{gonzales-lopez20} and \citet[][GOODS-ALMA survey]{gomez-guijarro21}, to number counts from an ALMA protocluster field with removed cluster galaxies from \citet{umehata17}, to the number counts from ALMA calibrator pointings from \citet{chen23}, and to number counts from single-dish blind surveys from \citet{lindner11}, \citet{scott12} and \citet{bing23}. 
There is an overall very good agreement between our number counts and these previous studies. 
Our number counts bridge the flux density range between the deepest ALMA blind survey and the deepest single-dish surveys, reaching nearly as deep as ASPECS-LP but covering a wider area than any individual ALMA blind survey.
In the bright end of Band 6, our number counts favour the results from the single-dish surveys over the ALMA-based results from \citet{gomez-guijarro21}, the latter likely being affected by large cosmic variance due to a small areal coverage.
In the faint end, our number counts agree with those from \citet{gonzales-lopez20}, although we do not reach quite the same depth as ASPECS-LP. 
This difference in depth is attributed to the higher S/N$_{\rm peak}$ source extraction threshold chosen in our work \citep[S/N$_{\rm peak}=5.4$ instead of 3.3 in][]{gonzales-lopez20}.

Band 7 is the only band in which only single pointings and no blind surveys are available in our database. 
We compare our results with ALMA-based number counts from \citet{oteo16}, \citet{bethermin20}, and \citet{chen23}, who all used single pointings with masked-out targets, and from \citet{stach18} and \citet{simpson20}, which are based on single-dish follow-up studies.
The number counts are largely consistent with the literature within the error bars. 
Our number counts are systematically a bit lower than the ones from \citet{bethermin20} that are based on their full sample, only masking a circular region of radius 1\,arcsec around the phase centre of the ALPINE pointings.
However, our number counts are in almost perfect agreement with those inferred from their `secure' sample which is furthermore limited to sources at $z<4$ with an optical or near-IR counterpart.
Even though this `secure' sample is meant to provide a lower limit, the resulting number counts are not systematically lower than ours.
Unfortunately, the \acosmos database does not yet probe the bright end in Band 7 as far as the single-dish follow-up studies, because we do not probe large enough areas to serendipitously detect those bright sources. 

For sanity check, we also report the number counts separately for blind surveys and single pointings (see Appendix~\ref{appendix_split}). 
We find that both results are consistent within the uncertainties in their common flux density ranges. 
This confirms the practical applicability of our blinding method for single pointings.

\subsection{Contribution to the cosmic infrared background}

\begin{table}
  \caption[]{Fraction of the CIB resolved into individual sources in our \acosmos and \agoodss databases.}
     \label{table-cib}
 $$ 
     \begin{tabular}{lcccc}
        \hline
        \noalign{\smallskip}
        \makecell[l]{ALMA Band} & \makecell[c]{$S_{\rm total, ALMA}$ \\ {[}Jy deg$^{-2}${]} } & \makecell[c]{$S_{\rm total, COBE}$ \\ {[}Jy deg$^{-2}${]} } & \makecell[c]{Percentage \\ resolved}  \\
        \noalign{\smallskip}
        \hline
        \noalign{\smallskip}
        Band 3$^{*)}$       &   $0.08\substack{+0.02\\-0.02}$       &  $2.2\pm1.1$            &   $4\substack{+2\\-2}$     \\
        Band 4              &   $1.0\substack{+0.6\\-0.3}$          &  $5.5\pm2.6$            &   $18\substack{+14\\-10}$  \\
        Band 6              &   $ 6.6\substack{+0.5\\-0.3}$         &  $19.1\pm8.2$           &   $34\substack{+15\\-15}$  \\
        Band 7              &   $ 14.5\substack{+5.6\\-2.1}$        &  $36.3\pm14.8$  &   $40\substack{+22\\-17} $ \\
       
        \noalign{\smallskip}
        \hline
     \end{tabular}
 $$ 
    Notes: Total flux density per unit area inferred from \acosmos and \agoodss number counts and CIB emission measured with \textit{COBE} \citep[][]{fixsen98} for ALMA Bands 3 ($2998\mu$m), 4 ($2082\mu$m), 6 ($1233\mu$m), and 7 ($925\mu$m). 
    The last column gives the fraction of the CIB that is resolved into discrete sources using our number counts. 
    
    $^{*)}$ Band 3 estimates for \textit{COBE} are an extrapolation as they lie outside of the data range constrained in \citet{fixsen98}.
 
\end{table}

Using our number counts, we can calculate the total flux density per unit area as emitted by these sources with S/N$_{\rm peak}\geq~5.4$ for the ALMA Bands 3, 4, 6, and 7. 
This is done by summing over the redshift-weighted (see Sect.~\ref{subsubsec_pairs}) total flux densities of all sources in each band divided by the respective effective area:

\begin{equation}
    S_{\rm total, ALMA} = \sum_i \frac{S_{\rm i} \cdot (1 - P_{\textup{pair},i})}{A_{\textup{eff}}(S_i, \theta_i)}. \label{eq_cib_contribution_weighted}
\end{equation}
This allows us to determine how much of the cosmic infrared background (CIB) is resolved into discrete sources at each wavelength. 
As a reference for the intensity of the CIB, we took the analytical fit by \citet{fixsen98} to the far-IR background light as measured by \textit{COBE}/FIRAS observations. 
This relation was fitted on the wavelength range between 125\,$\mu$m and 2000\,$\mu$m, which covers our Bands 4, 6, and 7. 
Even though Band 3 at $\sim$3\,mm is outside of this range, we still used the \citet{fixsen98} fit as a comparative value, although we emphasise that this is an ex\-tra\-po\-la\-tion. 
The total flux densities per unit area emitted by all sources detected in a given band ($S_{\rm total, ALMA}$) as well as the ones inferred from the \textit{COBE} data ($S_{\rm total, COBE}$) are listed in Table~\ref{table-cib}, in addition to the resulting fraction of the CIB that is thus resolved into discrete ALMA sources.

The percentage of the CIB resolved decreases notably with the increase in wavelength.
While in Band 7, we resolve $\sim$40\% of the CIB, in Band 6 it is $\sim$34\% and in Band 4 only $\sim$18\%. 
From our extrapolation in Band 3, the resolved fraction is even below 10\%.
We note, however, that in all cases the uncertainties are large, not least due to the already large uncertainties on the \citet{fixsen98} fit.
This decline with longer wavelengths is probably attributed to the fact that ALMA is mainly probing the Rayleigh-Jeans tail of the dust emission of DSFGs, which gets progressively fainter going further into the millimetre regime. 
Although the $\sim$2-3\,mm wavelength regime serves as a filter for very high-redshift SFGs \citep[][]{casey18b,casey18,casey21}, detecting the faint emission of these DSFGs in the millimetre still requires very long integration times, even with the good sensitivity of ALMA. 
Therefore, at these long wavelengths, we predominantly detect the brighter end of the population of DSFGs, missing out on most of the fainter ones which constitute the bulk of the total emission.
Our sample is thus not representative of the majority of this population, so there is still a lot of information missing on their dust emissivity and general properties.
Future observations with instruments of even higher sensitivity than ALMA at millimetre and longer wavelengths, for example the planned `Next Generation VLA’, will also aid in improving our understanding of this population by detecting more of these faint galaxies.

The fraction of the CIB resolved by ALMA into individual sources in Band 7 is higher than typical single-dish surveys achieve at similar wavelengths \citep[$\sim$20-30\%, e.g.][]{eales99,coppin06}, but consistent within the errors with results from \citet{smail02} where they resolve $60\%$ using lensed sources.
In Band 6, our resolved fraction is higher than single-dish results at 1.1\,mm from \citet{scott10} and \citet{hatsukade11} who both resolve around 6-10\%, but in agreement with \citet{viero13} who find $45\pm8\%$ via a stacking analysis.

Not only is ALMA already largely outperforming single-dish observations in resolving the CIB into individual sources, but it is currently only limited by integration time and can thus potentially resolve even higher fractions by performing deeper observations. 
Single-dish telescopes provide precise measurements of the brighter populations of DSFGs, but are already mostly at their confusion limits.
These limits are impossible to circumvent outside of using lensed sources.

\subsection{Comparison with number counts models}

\begin{figure*}[ht!]
\centering
\includegraphics{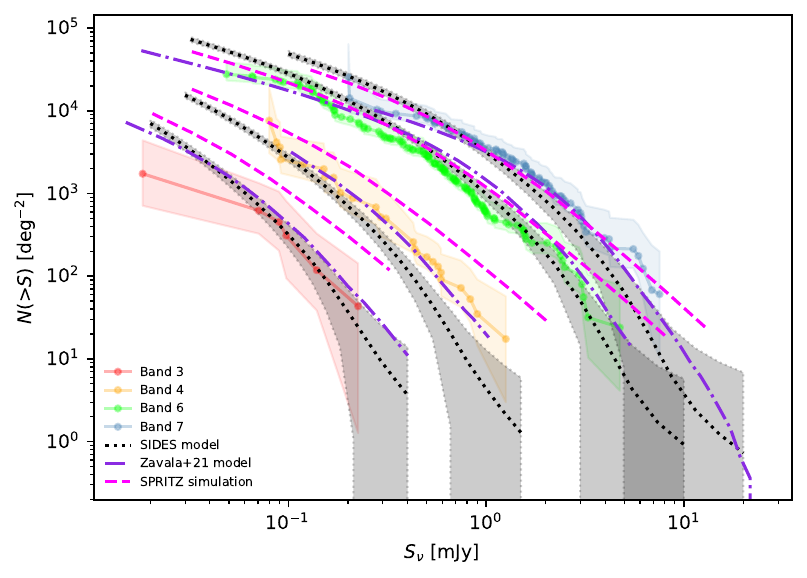}
\caption{Cumulative number counts as shown in Fig. \ref{figure-nc-all} in comparison with the models from \citet{zavala21}, \citet{bisigello21} (i.e. the SPRITZ simulation), and \citet{gkogkou23} (i.e. the SIDES model). The shaded regions of the SIDES curves indicate one standard deviation due to cosmic variance for an area corresponding to our coverage in each band.}
\label{counts_with_model}
\end{figure*}

We plot our number counts in comparison to the models from \citet{zavala21}, from the SPRITZ simulation of \citet{bisigello21}, and from the SIDES simulation of \citet{gkogkou23}. 
The \citet{zavala21} model is based on galaxy dust SEDs that were combined with an IR luminosity function (IRLF) with some free parameters to create mock sky maps matching the number counts at 1.2\,mm, 2\,mm, and 3\,mm available at that time.
The SPRITZ simulation \citep{bisigello21} is based on a set of IRLFs for different galaxy populations and their corresponding semi-empirical SED templates from which physical galaxy properties and fluxes were derived to create simulated catalogues and images.
The SIDES simulation \citep[][see also \citealt{bethermin17}]{gkogkou23} was built from dark matter lightcones that were populated with galaxies based on a stellar mass function using an abundance matching technique. 
These galaxies were randomly split into star-forming and passive populations based on a stellar mass- and redshift-dependent probability, derived from the observed evolution of the star-forming fraction \citep{davidzon17}.
They were subsequently assigned SFRs and SEDs based on their stellar mass and type (either main sequence galaxies or starbursts) to infer their observable flux densities in several filters. 
The model also considers lensing (de-)magnification and clustering properties.
The comparison of the models with our measured number counts is shown in Fig.~\ref{counts_with_model}. 

The predictions from the \citet{zavala21} model agree well with our measured number counts, thus their galaxy evolution model provides an overall good representation of the (sub)millimetre galaxy population.
This is not entirely surprising as this phenomenological model was developed to reproduce the number count from real 1.2\,mm, 2\,mm, and 3\,mm data available at that time. 
Nevertheless, there are notable differences in Band 6 in the intermediate flux density range (i.e. 0.2-2.0\,mJy) and Band 7 in the high flux density range (>4\,mJy).
These differences are probably due to the limited number counts available at that time used for fitting the model in \citet{zavala21}. 
In particular, in Band~6, the only constraints used were the ASPECS-LP number counts below $\sim$0.4\,mJy. Higher flux densities in Band 6 or shorter wavelengths were not constrained with observational data.
Despite these slight differences, the model of \citet{zavala21} provides a reasonable representation of the (sub)millimetre galaxy population. 
It is noteworthy that predictions at low flux density in Band~4, where no constraints were available at the time of creation of this model, are fully consistent with our measurements. 
This highlights the predictive power of this model.

The predicted number counts from SPRITZ are in good agreement with our measurements in Band~7.
Although consistent with our number counts within the uncertainties in Band 6 and the faint end of Band 4, SPRITZ predicts significantly higher number counts (by a factor $\sim$5-6) in the bright end of Band 4 and over the full range of Band 3.
The SPRITZ simulation and its underlying SED templates are derived from \textit{Herschel}-based optical-to-IR \citep[up to $\sim$500\,$\mu$m;][]{gruppioni13} data and optimised to make predictions in this specific regime.
The (sub)millimetre regime, however, was not explicitly taken into account in the construction of the simulation and thus it is not well constrained at significantly longer wavelengths.

The SIDES model agrees with our measured number counts within the uncertainties. 
This is remarkable considering that SIDES was not explicitly fitted to any measured (sub)millimetre number counts.
We notice, however, that the model tends to predict systematically lower number counts at the bright end of Bands~4 and 7. 
A slight under-prediction of SIDES at the bright end of the galaxy number counts was previously noted by \citet{bing23}, who suggested a lack of either cold dust in the underlying SEDs or of galaxies with high SFRs.

\section{Summary} \label{sec_summary}

In this work we presented the updated \acosmos database, including the newly added \agoodss, in data version \texttt{20220606}. 
This more than doubles the size of the \acosmos database compared to the previous release. 
With a robust sample of $\sim$1,400 galaxies over a wide redshift range and with SED-inferred information such as SFR and gas or dust masses, it is a rich resource for studies on (dusty) galaxy evolution. 
This database, which includes the continuum images, blind and prior photometry catalogues,
and value-added galaxy catalogues, is made public to the community and will receive regular updates.

We used the blind continuum detections from this database to infer in a homogeneous way galaxy number counts across all major ALMA bands. 
To reduce the inhomogeneous observational bias inherent to the \acosmos and \agoodss databases, we used a conservative combination of selections and corrections: 
we restricted our a\-na\-ly\-sis to blind surveys or single targeted pointings with a secure target at their phase centre, masked a circular area around the phase centre of these single pointings, and weighted any source detected outside of this mask by redshift proximity to the target. 
Finally, we used simulations based on the SIDES model \citep[][]{bethermin17,gkogkou23}, to confirm the applicability of our method. 

Thanks to the sensitivity of ALMA and the large areal coverage offered by the combination of all ALMA archive projects in the COSMOS and GOODS-S fields, the number counts inferred here bridge the flux density gap between single-dish observations, which constrain only the bright end of the number counts, and pencil beam ALMA blind surveys, which constrain the faint end of the number counts (albeit with large uncertainties due to cosmic variance).
In the flux density ranges in common, our number counts are overall in good agreement with those retrieved from the literature. 
Finally, down to the flux density limits of our number counts, about 40\%, 34\%, 18\%, and 4\% of the CIB is resolved into discrete sources in Band~7, 6, 4, and 3, respectively. 
Thus, while the emission of DSFGs is well constrained at 0.9-1.3\,mm wavelengths, it remains poorly understood at longer wavelengths. 
However, as ALMA observations are far less limited by confusion noise than single-dish facilities, future deeper observations with ALMA will undoubtedly allow us to resolve an ever larger fraction of the CIB at these wavelengths.

The homogeneous measurements of the number counts in ALMA Bands~3, 4, 6, and 7 over such broad flux density ranges offer unique constraints for galaxy evolution models. 
We thus compared our number counts with predictions from the models of \citet{zavala21}, SPRITZ \citep{bisigello21}, and SIDES \citep[][]{bethermin17,gkogkou23}. 
We find that our results are in very good agreement with the predictions from the model of \citet{zavala21}. 
The SPRITZ model is consistent with our number counts in Bands~6 and 7, but its predictions are significantly higher than the observations at longer wavelenghts (i.e. Bands~3 and 4).
Our number counts are overall in good agreement with those predicted by the SIDES simulation, even thought this one was not explicitly fitted to any measured (sub)millimetre number counts. 
However, this simulation tends to slightly underestimate the number of bright sources in Bands~4 and 7.   

We have shown that it is possible to infer number counts from the heterogeneous target-driven \acosmos database. 
This method naturally has limitations; however, the effects on the number counts are negligible compared to the effects of cosmic variance.   
Our analysis therefore profits from the large areal coverage of the \acosmos database and the availability of blind surveys within it. 
Future updates will allow us to decrease the cosmic variance and further improve the accuracy of the measured number counts.

\begin{acknowledgements}
SA would like to thank Laura Bisigello and Jorge Zavala for providing the curves of their respective models.
FB and SA gratefully acknowledge the Collaborative Research Center 1601 (SFB 1601 sub-project C2) funded by the Deutsche Forschungsgemeinschaft (DFG, German Research Foundation) – 500700252.
ID acknowledges support from INAF Minigrant `Harnessing the power of VLBA towards a census of AGN and star formation at high redshift'.

This paper makes use of the following ALMA data: 
\texttt{ADS/JAO.ALMA\#2011.0.00064.S}, \texttt{ADS/JAO.ALMA\#2011.0.00097.S}, \texttt{ADS/JAO.ALMA\#2011.0.00539.S}, \texttt{ADS/JAO.ALMA\#2011.0.00742.S}, \texttt{ADS/JAO.ALMA\#2012.1.00076.S}, \texttt{ADS/JAO.ALMA\#2012.1.00173.S}, \texttt{ADS/JAO.ALMA\#2012.1.00307.S}, \texttt{ADS/JAO.ALMA\#2012.1.00323.S}, \texttt{ADS/JAO.ALMA\#2012.1.00523.S}, \texttt{ADS/JAO.ALMA\#2012.1.00536.S}, \texttt{ADS/JAO.ALMA\#2012.1.00775.S}, \texttt{ADS/JAO.ALMA\#2012.1.00869.S}, \texttt{ADS/JAO.ALMA\#2012.1.00919.S}, \texttt{ADS/JAO.ALMA\#2012.1.00952.S}, \texttt{ADS/JAO.ALMA\#2012.1.00978.S}, \texttt{ADS/JAO.ALMA\#2012.1.00983.S}, \texttt{ADS/JAO.ALMA\#2013.1.00034.S}, \texttt{ADS/JAO.ALMA\#2013.1.00092.S}, \texttt{ADS/JAO.ALMA\#2013.1.00118.S}, \texttt{ADS/JAO.ALMA\#2013.1.00139.S}, \texttt{ADS/JAO.ALMA\#2013.1.00146.S}, \texttt{ADS/JAO.ALMA\#2013.1.00151.S}, \texttt{ADS/JAO.ALMA\#2013.1.00171.S}, \texttt{ADS/JAO.ALMA\#2013.1.00205.S}, \texttt{ADS/JAO.ALMA\#2013.1.00208.S}, \texttt{ADS/JAO.ALMA\#2013.1.00250.S}, \texttt{ADS/JAO.ALMA\#2013.1.00276.S}, \texttt{ADS/JAO.ALMA\#2013.1.00470.S}, \texttt{ADS/JAO.ALMA\#2013.1.00668.S}, \texttt{ADS/JAO.ALMA\#2013.1.00718.S}, \texttt{ADS/JAO.ALMA\#2013.1.00786.S}, \texttt{ADS/JAO.ALMA\#2013.1.00815.S}, \texttt{ADS/JAO.ALMA\#2013.1.00836.S}, \texttt{ADS/JAO.ALMA\#2013.1.00884.S}, \texttt{ADS/JAO.ALMA\#2013.1.00914.S}, \texttt{ADS/JAO.ALMA\#2013.1.01258.S}, \texttt{ADS/JAO.ALMA\#2013.1.01271.S}, \texttt{ADS/JAO.ALMA\#2013.1.01292.S}, \texttt{ADS/JAO.ALMA\#2015.1.00026.S}, \texttt{ADS/JAO.ALMA\#2015.1.00039.S}, \texttt{ADS/JAO.ALMA\#2015.1.00040.S}, \texttt{ADS/JAO.ALMA\#2015.1.00055.S}, \texttt{ADS/JAO.ALMA\#2015.1.00098.S}, \texttt{ADS/JAO.ALMA\#2015.1.00122.S}, \texttt{ADS/JAO.ALMA\#2015.1.00137.S}, \texttt{ADS/JAO.ALMA\#2015.1.00207.S}, \texttt{ADS/JAO.ALMA\#2015.1.00220.S}, \texttt{ADS/JAO.ALMA\#2015.1.00228.S}, \texttt{ADS/JAO.ALMA\#2015.1.00260.S}, \texttt{ADS/JAO.ALMA\#2015.1.00299.S}, \texttt{ADS/JAO.ALMA\#2015.1.00379.S}, \texttt{ADS/JAO.ALMA\#2015.1.00388.S}, \texttt{ADS/JAO.ALMA\#2015.1.00540.S}, \texttt{ADS/JAO.ALMA\#2015.1.00543.S}, \texttt{ADS/JAO.ALMA\#2015.1.00568.S}, \texttt{ADS/JAO.ALMA\#2015.1.00664.S}, \texttt{ADS/JAO.ALMA\#2015.1.00704.S}, \texttt{ADS/JAO.ALMA\#2015.1.00821.S}, \texttt{ADS/JAO.ALMA\#2015.1.00853.S}, \texttt{ADS/JAO.ALMA\#2015.1.00861.S}, \texttt{ADS/JAO.ALMA\#2015.1.00862.S}, \texttt{ADS/JAO.ALMA\#2015.1.00870.S}, \texttt{ADS/JAO.ALMA\#2015.1.00907.S}, \texttt{ADS/JAO.ALMA\#2015.1.00928.S}, \texttt{ADS/JAO.ALMA\#2015.1.00948.S}, \texttt{ADS/JAO.ALMA\#2015.1.01074.S}, \texttt{ADS/JAO.ALMA\#2015.1.01096.S}, \texttt{ADS/JAO.ALMA\#2015.1.01105.S}, \texttt{ADS/JAO.ALMA\#2015.1.01111.S}, \texttt{ADS/JAO.ALMA\#2015.1.01171.S}, \texttt{ADS/JAO.ALMA\#2015.1.01205.S}, \texttt{ADS/JAO.ALMA\#2015.1.01212.S}, \texttt{ADS/JAO.ALMA\#2015.1.01362.S}, \texttt{ADS/JAO.ALMA\#2015.1.01379.S}, \texttt{ADS/JAO.ALMA\#2015.1.01447.S}, \texttt{ADS/JAO.ALMA\#2015.1.01495.S}, \texttt{ADS/JAO.ALMA\#2015.1.01590.S}, \texttt{ADS/JAO.ALMA\#2015.A.00009.S}, \texttt{ADS/JAO.ALMA\#2015.A.00026.S}, \texttt{ADS/JAO.ALMA\#2016.1.00012.S}, \texttt{ADS/JAO.ALMA\#2016.1.00048.S}, \texttt{ADS/JAO.ALMA\#2016.1.00171.S}, \texttt{ADS/JAO.ALMA\#2016.1.00279.S}, \texttt{ADS/JAO.ALMA\#2016.1.00324.L}, \texttt{ADS/JAO.ALMA\#2016.1.00330.S}, \texttt{ADS/JAO.ALMA\#2016.1.00463.S}, \texttt{ADS/JAO.ALMA\#2016.1.00478.S}, \texttt{ADS/JAO.ALMA\#2016.1.00534.S}, \texttt{ADS/JAO.ALMA\#2016.1.00564.S}, \texttt{ADS/JAO.ALMA\#2016.1.00567.S}, \texttt{ADS/JAO.ALMA\#2016.1.00624.S}, \texttt{ADS/JAO.ALMA\#2016.1.00646.S}, \texttt{ADS/JAO.ALMA\#2016.1.00721.S}, \texttt{ADS/JAO.ALMA\#2016.1.00726.S}, \texttt{ADS/JAO.ALMA\#2016.1.00735.S}, \texttt{ADS/JAO.ALMA\#2016.1.00778.S}, \texttt{ADS/JAO.ALMA\#2016.1.00790.S}, \texttt{ADS/JAO.ALMA\#2016.1.00798.S}, \texttt{ADS/JAO.ALMA\#2016.1.00804.S}, \texttt{ADS/JAO.ALMA\#2016.1.00932.S}, \texttt{ADS/JAO.ALMA\#2016.1.00954.S}, \texttt{ADS/JAO.ALMA\#2016.1.00967.S}, \texttt{ADS/JAO.ALMA\#2016.1.00990.S}, \texttt{ADS/JAO.ALMA\#2016.1.01001.S}, \texttt{ADS/JAO.ALMA\#2016.1.01012.S}, \texttt{ADS/JAO.ALMA\#2016.1.01040.S}, \texttt{ADS/JAO.ALMA\#2016.1.01079.S}, \texttt{ADS/JAO.ALMA\#2016.1.01155.S}, \texttt{ADS/JAO.ALMA\#2016.1.01184.S}, \texttt{ADS/JAO.ALMA\#2016.1.01208.S}, \texttt{ADS/JAO.ALMA\#2016.1.01240.S}, \texttt{ADS/JAO.ALMA\#2016.1.01355.S}, \texttt{ADS/JAO.ALMA\#2016.1.01426.S}, \texttt{ADS/JAO.ALMA\#2016.1.01454.S}, \texttt{ADS/JAO.ALMA\#2016.1.01512.S}, \texttt{ADS/JAO.ALMA\#2016.1.01546.S}, \texttt{ADS/JAO.ALMA\#2016.1.01559.S}, \texttt{ADS/JAO.ALMA\#2016.1.01604.S}, \texttt{ADS/JAO.ALMA\#2017.1.00001.S}, \texttt{ADS/JAO.ALMA\#2017.1.00046.S}, \texttt{ADS/JAO.ALMA\#2017.1.00127.S}, \texttt{ADS/JAO.ALMA\#2017.1.00138.S}, \texttt{ADS/JAO.ALMA\#2017.1.00190.S}, \texttt{ADS/JAO.ALMA\#2017.1.00270.S}, \texttt{ADS/JAO.ALMA\#2017.1.00300.S}, \texttt{ADS/JAO.ALMA\#2017.1.00326.S}, \texttt{ADS/JAO.ALMA\#2017.1.00373.S}, \texttt{ADS/JAO.ALMA\#2017.1.00413.S}, \texttt{ADS/JAO.ALMA\#2017.1.00428.L}, \texttt{ADS/JAO.ALMA\#2017.1.00486.S}, \texttt{ADS/JAO.ALMA\#2017.1.00604.S}, \texttt{ADS/JAO.ALMA\#2017.1.00755.S}, \texttt{ADS/JAO.ALMA\#2017.1.00856.S}, \texttt{ADS/JAO.ALMA\#2017.1.00893.S}, \texttt{ADS/JAO.ALMA\#2017.1.01020.S}, \texttt{ADS/JAO.ALMA\#2017.1.01027.S}, \texttt{ADS/JAO.ALMA\#2017.1.01099.S}, \texttt{ADS/JAO.ALMA\#2017.1.01163.S}, \texttt{ADS/JAO.ALMA\#2017.1.01176.S}, \texttt{ADS/JAO.ALMA\#2017.1.01217.S}, \texttt{ADS/JAO.ALMA\#2017.1.01259.S}, \texttt{ADS/JAO.ALMA\#2017.1.01276.S}, \texttt{ADS/JAO.ALMA\#2017.1.01347.S}, \texttt{ADS/JAO.ALMA\#2017.1.01358.S}, \texttt{ADS/JAO.ALMA\#2017.1.01359.S}, \texttt{ADS/JAO.ALMA\#2017.1.01451.S}, \texttt{ADS/JAO.ALMA\#2017.1.01471.S}, \texttt{ADS/JAO.ALMA\#2017.1.01512.S}, \texttt{ADS/JAO.ALMA\#2017.1.01618.S}, \texttt{ADS/JAO.ALMA\#2017.1.01659.S}, \texttt{ADS/JAO.ALMA\#2017.1.01674.S}, \texttt{ADS/JAO.ALMA\#2017.1.01677.S}, \texttt{ADS/JAO.ALMA\#2017.1.01713.S}, \texttt{ADS/JAO.ALMA\#2017.A.00013.S}, \texttt{ADS/JAO.ALMA\#2017.A.00032.S}, \texttt{ADS/JAO.ALMA\#2017.A.00034.S}, \texttt{ADS/JAO.ALMA\#2018.1.00081.S}, \texttt{ADS/JAO.ALMA\#2018.1.00085.S}, \texttt{ADS/JAO.ALMA\#2018.1.00164.S}, \texttt{ADS/JAO.ALMA\#2018.1.00216.S}, \texttt{ADS/JAO.ALMA\#2018.1.00222.S}, \texttt{ADS/JAO.ALMA\#2018.1.00231.S}, \texttt{ADS/JAO.ALMA\#2018.1.00236.S}, \texttt{ADS/JAO.ALMA\#2018.1.00251.S}, \texttt{ADS/JAO.ALMA\#2018.1.00329.S}, \texttt{ADS/JAO.ALMA\#2018.1.00348.S}, \texttt{ADS/JAO.ALMA\#2018.1.00429.S}, \texttt{ADS/JAO.ALMA\#2018.1.00478.S}, \texttt{ADS/JAO.ALMA\#2018.1.00543.S}, \texttt{ADS/JAO.ALMA\#2018.1.00567.S}, \texttt{ADS/JAO.ALMA\#2018.1.00570.S}, \texttt{ADS/JAO.ALMA\#2018.1.00635.S}, \texttt{ADS/JAO.ALMA\#2018.1.00681.S}, \texttt{ADS/JAO.ALMA\#2018.1.00874.S}, \texttt{ADS/JAO.ALMA\#2018.1.00876.S}, \texttt{ADS/JAO.ALMA\#2018.1.00933.S}, \texttt{ADS/JAO.ALMA\#2018.1.00938.S}, \texttt{ADS/JAO.ALMA\#2018.1.00992.S}, \texttt{ADS/JAO.ALMA\#2018.1.01044.S}, \texttt{ADS/JAO.ALMA\#2018.1.01079.S}, \texttt{ADS/JAO.ALMA\#2018.1.01128.S}, \texttt{ADS/JAO.ALMA\#2018.1.01136.S}, \texttt{ADS/JAO.ALMA\#2018.1.01225.S}, \texttt{ADS/JAO.ALMA\#2018.1.01281.S}, \texttt{ADS/JAO.ALMA\#2018.1.01359.S}, \texttt{ADS/JAO.ALMA\#2018.1.01521.S}, \texttt{ADS/JAO.ALMA\#2018.1.01536.S}, \texttt{ADS/JAO.ALMA\#2018.1.01551.S}, \texttt{ADS/JAO.ALMA\#2018.1.01594.S}, \texttt{ADS/JAO.ALMA\#2018.1.01605.S}, \texttt{ADS/JAO.ALMA\#2018.1.01673.S}, \texttt{ADS/JAO.ALMA\#2018.1.01739.S}, \texttt{ADS/JAO.ALMA\#2018.1.01824.S}, \texttt{ADS/JAO.ALMA\#2018.1.01841.S}, \texttt{ADS/JAO.ALMA\#2018.1.01852.S}, \texttt{ADS/JAO.ALMA\#2018.1.01871.S}, \texttt{ADS/JAO.ALMA\#2018.A.00022.S}, \texttt{ADS/JAO.ALMA\#2018.A.00037.S}, \texttt{ADS/JAO.ALMA\#2019.1.00074.S}, \texttt{ADS/JAO.ALMA\#2019.1.00102.S}, \texttt{ADS/JAO.ALMA\#2019.1.00151.S}, \texttt{ADS/JAO.ALMA\#2019.1.00397.S}, \texttt{ADS/JAO.ALMA\#2019.1.00399.S}, \texttt{ADS/JAO.ALMA\#2019.1.00459.S}, \texttt{ADS/JAO.ALMA\#2019.1.00477.S}, \texttt{ADS/JAO.ALMA\#2019.1.00652.S}, \texttt{ADS/JAO.ALMA\#2019.1.00678.S}, \texttt{ADS/JAO.ALMA\#2019.1.00702.S}, \texttt{ADS/JAO.ALMA\#2019.1.00909.S}, \texttt{ADS/JAO.ALMA\#2019.1.00964.S}, \texttt{ADS/JAO.ALMA\#2019.1.01075.S}, \texttt{ADS/JAO.ALMA\#2019.1.01127.S}, \texttt{ADS/JAO.ALMA\#2019.1.01142.S}, \texttt{ADS/JAO.ALMA\#2019.1.01201.S}, \texttt{ADS/JAO.ALMA\#2019.1.01286.S}, \texttt{ADS/JAO.ALMA\#2019.1.01329.S}, \texttt{ADS/JAO.ALMA\#2019.1.01491.S}, \texttt{ADS/JAO.ALMA\#2019.1.01524.S}, \texttt{ADS/JAO.ALMA\#2019.1.01528.S}, \texttt{ADS/JAO.ALMA\#2019.1.01537.S}, \texttt{ADS/JAO.ALMA\#2019.1.01600.S}, \texttt{ADS/JAO.ALMA\#2019.1.01615.S}, \texttt{ADS/JAO.ALMA\#2019.1.01634.L}, \texttt{ADS/JAO.ALMA\#2019.1.01702.S}, \texttt{ADS/JAO.ALMA\#2019.1.01722.S}, \texttt{ADS/JAO.ALMA\#2019.1.01832.S}, \texttt{ADS/JAO.ALMA\#2019.2.00118.S}, \texttt{ADS/JAO.ALMA\#2019.2.00143.S}, \texttt{ADS/JAO.ALMA\#2019.2.00246.S}, \texttt{ADS/JAO.ALMA\#2019.A.00015.S}.
ALMA is a partnership of ESO (representing its member states), NSF (USA) and NINS (Japan), together with NRC (Canada), MOST and ASIAA (Taiwan), and KASI (Republic of Korea), in cooperation with the Republic of Chile. The Joint ALMA Observatory is operated by ESO, AUI/NRAO and NAOJ. 

\end{acknowledgements}

\bibliography{mybib.bib}

\appendix

\section{Counterpart association}  \label{appendix_cpa}

Here, we provide an overview of the counterpart association method and the machine learning (ML) model applied in the process.
For a detailed description we refer the reader to the work of \citet[][their Sect.~4.2 and Appendix~D]{a3cosmos_1}.

To assess the accuracy of the association between priors and extracted ALMA sources, we made use of several ancillary images from optical and near-infrared instruments covering the COSMOS and GOODS-S fields, as listed at the end of Sects.~\ref{subsec_cosmos} and \ref{subsec_goodss}, respectively.
The following parameters were then measured for each ALMA source-counterpart pair:

\begin{itemize}
\item The angular separation between the positions of the ALMA source and the counterpart, normalised by the angular radius of the ALMA source (`Sep.');
\item The signal-to-noise ratio of the total flux of the ALMA source (S/N$_{\rm ALMA}$);
\item The signal-to-noise ratio measured in the counterpart images at the positions of the ALMA source (S/N$_{\rm S}$) and at the reference position of the counterpart source (S/N$_{\rm Ref.}$), determined via an aperture photometry approach;
\item The amount of extended emission in the counterpart images in-between the ALMA and counterpart source positions, determined by measuring the flux in a series of apertures placed between the respective positions.
\end{itemize}

To assess the reliability of the association based on these parameters, \citet{a3cosmos_1} trained a supervised ML model using the Python package \texttt{Scikit-learn} \citep{skykit-learn} with the Gaussian Process classifier.  
The training data for the model consisted of a \acosmos subsample of $\sim$600 randomly selected sources which had been manually classified as either robust or spurious counterpart. 
The accuracy of the trained model was then assessed by running it on a validation sample of another $\sim$400 manually classified sources not contained in the training data. 
They found an agreement between the manual and ML classification for $\sim$97\% of the sources in the validation sample. 
The same accuracy was also obtained for the combined sample of $\sim$1000 sources.

For our \acosmos database update, we applied this trained model to determine the reliability of our counterpart association for the newer COSMOS data, using the same counterpart images as \citet{a3cosmos_1}, only replacing the \textit{Spitzer}/IRAC 3.6\,$\mu$m image by a newer image from \citet{moneti22} (see Sect.~\ref{subsec_cosmos}). 
However, naturally, new counterpart images and most importantly different filters had to be used for GOODS-S, making the model not applicable to \agoodss.
The number of sources in our current version of the \agoodss prior source catalogue (i.e. 568) approximately corresponds to the size of the initial training dataset of the ML model.
We therefore classified all \agoodss sources manually, creating the training dataset for the ML model in preparation for the future updates of the \agoodss database.

\section{\acosmos and \agoodss sky coverage by band}  \label{appendix_coverage}

Here, we show the spatial sky coverage of our \acosmos and \agoodss databases, as displayed cumulatively in Fig.~\ref{sky_coverage}, separately for each ALMA band. The values of the individual coverages are also listed in Table~\ref{tab_database}.

\newpage

\begin{landscape}\centering

\vspace*{\fill}
\begin{figure}[htpb]
\centering
\includegraphics[width=\linewidth]{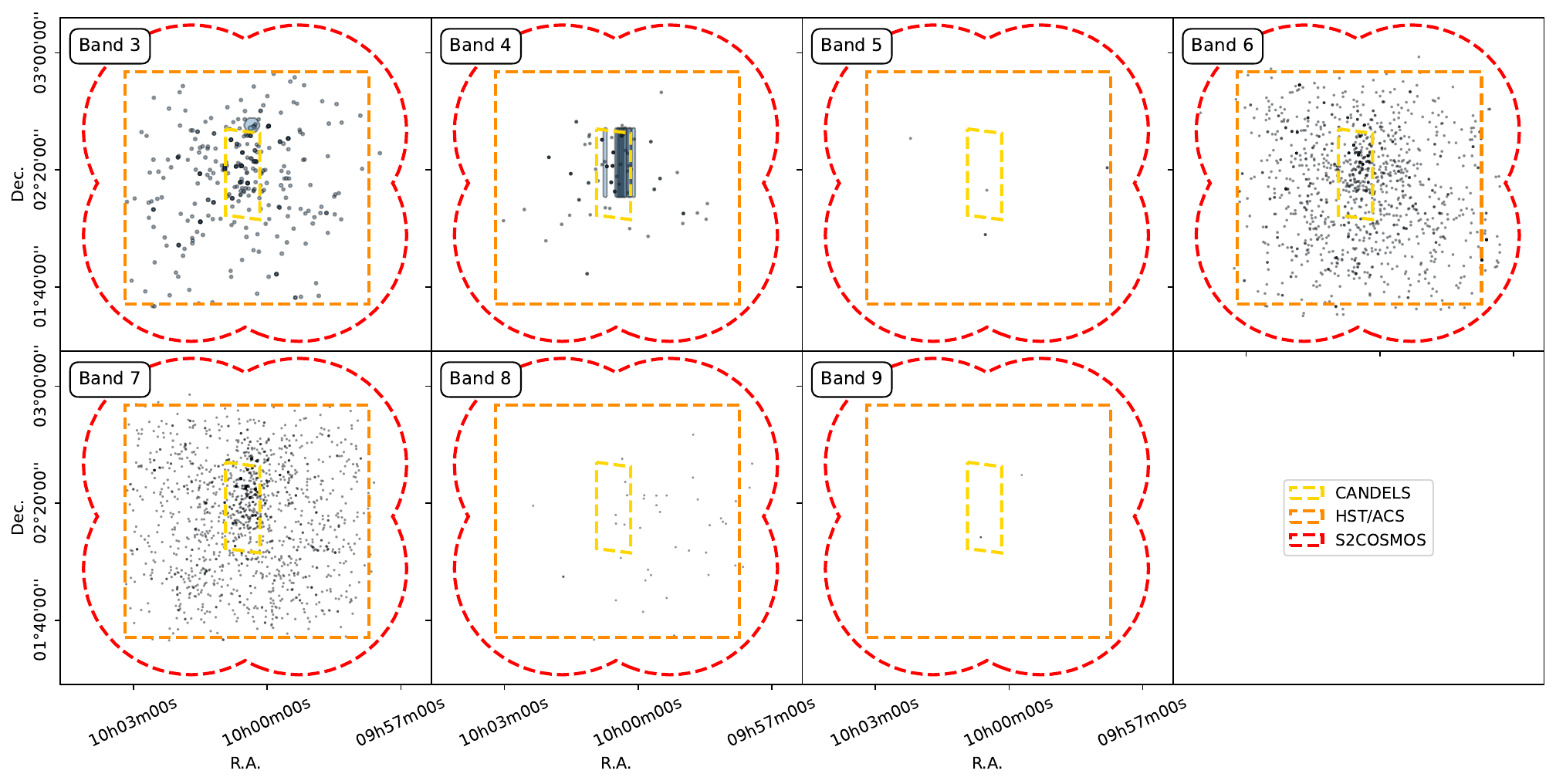}
\caption{Spatial sky coverage of \acosmos as shown in the top panel of Fig. \ref{sky_coverage}, and split into individual ALMA bands, as indicated in the top left of each panel. The dashed coloured lines indicate the approximate outlines of the CANDELS \citep[yellow;][]{grogin11}, COSMOS \textit{HST}/ACS \citep[orange;][]{koekemoer07}, and S2COSMOS \citep[red;][]{simpson19} fields.}
\label{a3cosmos_coverage_by_band}
\end{figure}
\vfill
\end{landscape}

\newpage

\begin{landscape}\centering
\vspace*{\fill}
\begin{figure}[htpb]
\centering
\includegraphics[width=\linewidth]{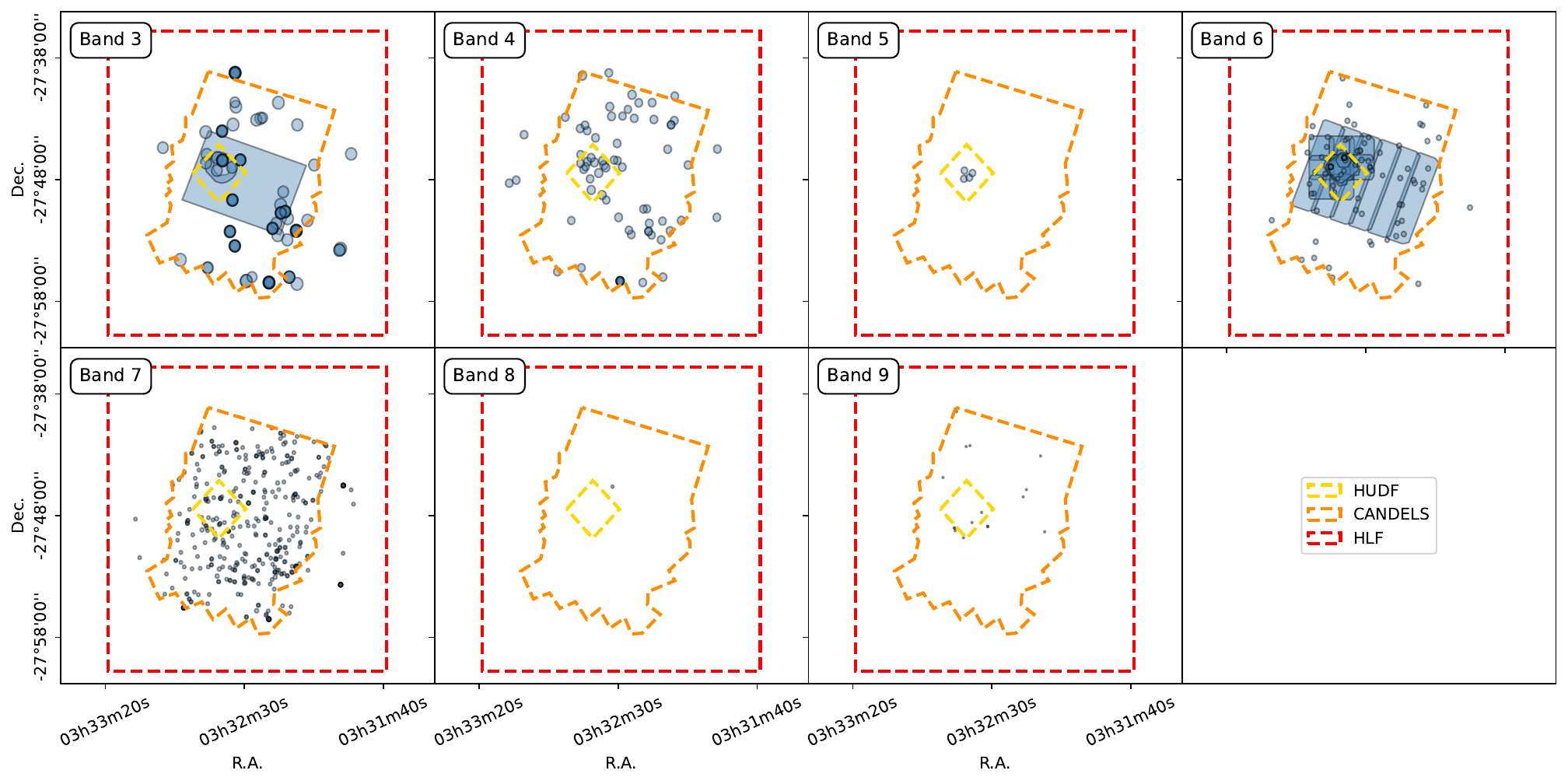}
\caption{Spatial sky coverage of \agoodss as shown in the bottom panel of Fig. \ref{sky_coverage}, and  split into individual ALMA bands, as indicated in the top left of each panel. The dashed coloured lines indicate the approximate outlines of the Hubble Ultra Deep Field \citep[yellow;][]{hudf}, CANDELS \citep[orange; as in][]{guo13}, and Hubble Legacy Fields \citep[red;][]{hlf}.}
\label{a3goodss_coverage_by_band}
\end{figure}
\vfill
\end{landscape}

\newpage
\section{Simulating observations in SIDES}  \label{appendix_sim}

In order to `observe' mock sources in each of our 58 realisations of a 2\,deg$^2$ sky field made from SIDES, we created a set of mock pointings. 
For this, we used the observational properties (i.e. areal coverage and sensitivity profile) of each precisely targeted pointing used in our real number counts analysis.
Since the SIDES simulation does not include source size information we, however, assumed an angular resolution of 1\,arcsec (which corresponds to the mean angular resolution in our \acosmos database) for all mock pointings and that all mock sources are point sources at this resolution.

\begin{figure}
\centering
\includegraphics[width=\linewidth]{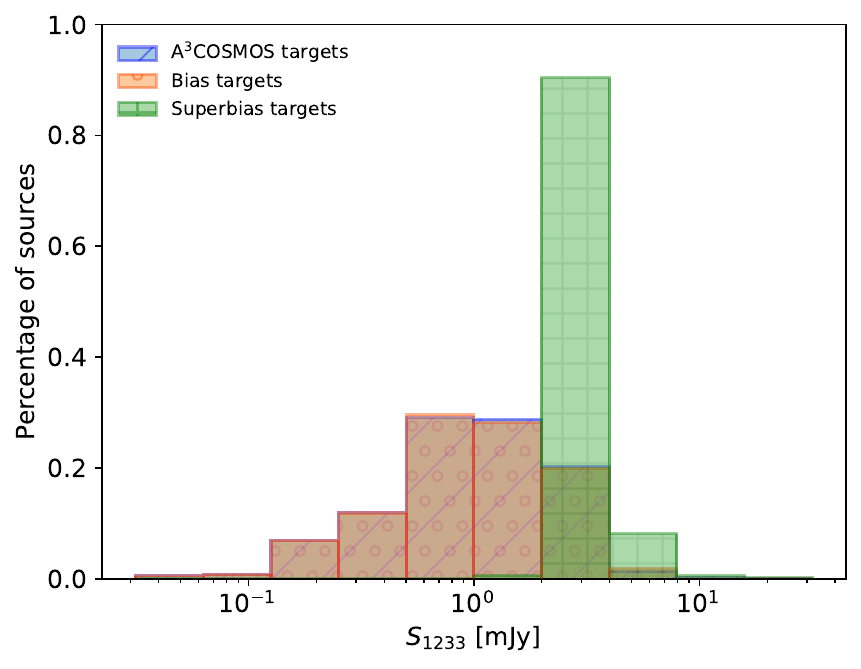}
\caption{Band 6 flux density distribution of targets in the simulation compared to the real targets of precisely targeted pointings in the \acosmos database. Orange with circles indicates  Bias mode; green with squares is  Superbias mode; and  blue with diagonal lines is for  real targets.}
\label{sim_target_flux_distribution}
\end{figure}

These mock pointings were then placed in each of the 58 fields, centred on mock target sources according to the respective observational bias mode (i.e. Bias, Superbias or Random, see Sect.~\ref{subsec_sim}).
We selected sources from the SIDES catalogue in each respective field as mock targets based on their flux density (except the Random mode in which the pointings were distributed randomly). 
In the Superbias mode, all the brightest sources in a given field were chosen as targets. 
In the Bias mode, mock target sources were drawn randomly based on their flux density, with a probability according to the flux density distribution of the real target sources in all precisely targeted pointings in our database in the ALMA band in consideration. 
Figure~\ref{sim_target_flux_distribution} shows the flux density distribution of our chosen mock targets in the Bias and Superbias modes, compared to that of the real \acosmos targets in Band 6. 
The Bias and real target distributions are almost identical by construction, with the majority of targets being in the intermediate to higher flux density range ($\sim$\,$0.5-4$\,mJy).
In contrast, Superbias targets are exclusively bright sources ($\gtrsim$\,2\,mJy).
In both the Bias and Superbias modes, the choice of which mock pointing to assign to which mock target was made to ensure that every mock target is detectable in its assigned mock pointing (as it is the case in our set of real pointings).

The source extraction from these mock observations followed the same constraint as our real source extraction, that is, only sources with S/N$_{\textup{peak}}>5.4$ were considered as detected, but assuming that above this threshold $C_{\textup{compl.}} = 1$ and $C_{\textup{contam.}} = 0$.  
Here, S/N$_{\textup{peak}}$ is calculated from the sensitivity of a mock pointing at the position of the mock source covered and the flux density of that source as given in the SIDES catalogue (i.e. $S_{\rm peak}=S_{\rm total}$, as all mock sources are assumed to be point sources).

For the corrected mock number counts, a central circular region around the phase centre of each mock pointing was disregarded during source extraction.
In addition, in order to allow for a weighting of mock sources by redshift proximity to their respective mock targets (as in our real analysis), we assigned to each detected mock source a mock redshift uncertainty.
For this, we randomly assigned each detected mock (serendipitous) source and each mock target to either a photometric, spectroscopic, or no redshift, based on the fractions of our real sources (see Table~\ref{tab_z_percentages}). 
We then assigned to each mock source and mock target a redshift randomly picked within a Gaussian PDF centred on their original SIDES redshift, $z$, with a width of $0.001(1+z)$ for spectroscopic or $0.06(1+z)$ for photometric redshifts. 
These assigned redshifts were then used for weighting as described in Sect.~\ref{subsubsec_pairs}.

\begin{table}
  \caption[]{Fractions of sources detected within precisely targeted \acosmos pointings that have a prior spectroscopic, photometric, or no prior redshift, split into sources within a radius of 1\,arcsec around the phase centre (Targets) and sources further away from the phase centre than 1\,arcsec (Serendipitous).}
     \label{tab_z_percentages}
 $$ 
     \begin{tabular}{lcc}
        \hline
        \noalign{\smallskip}
           &  Targets  &  Serendipitous  \\
        \noalign{\smallskip}
        \hline
        \noalign{\smallskip}
         Spec. $z$        &  40.66\% &    9.23\%    \\
         Photo. $z$       &  54.58\% &   73.85\%    \\
         No prior $z$     &  4.76\%  &   16.92\%    \\      
        \noalign{\smallskip}
        \hline
     \end{tabular}
 $$ 
\end{table}

\section{Selecting the radius of the central mask} \label{appendix_mask}

\begin{figure*}
\centering
\includegraphics[width=0.8\linewidth]{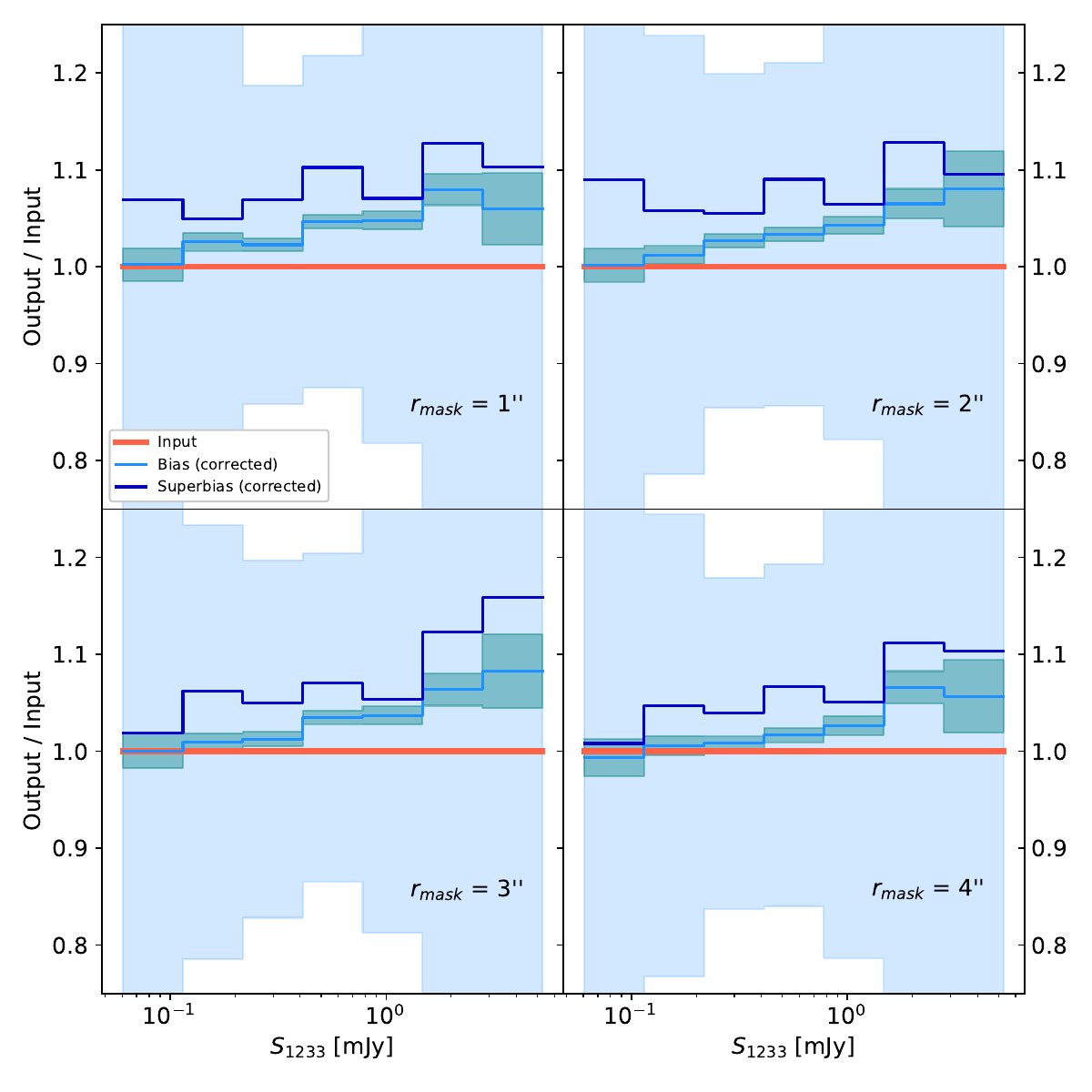}
\caption{Simulated differential number counts in ALMA Band 6, retrieved with the corrected Bias and Superbias modes for four different central mask radii, divided by the underlying Input number counts. Shown here is the mean over ten run-throughs of the simulation (i.e. 580 realisations). The light shaded regions show the standard deviation due to cosmic variance; the  dark shaded regions show the error on the mean (i.e. $\sigma_{\rm cos. var.}/\sqrt{580}$). The errors are of approximately the same size for the Bias and Superbias modes, but for clarity they are only shown for the Bias mode.}
\label{sim_mask_comp_diff}
\end{figure*}

We wanted to determine the optimal size of the mask put on the centre of our precisely targeted pointings. 
To do so, we ran our simulation (see Sect.~\ref{subsec_sim}) with several mask radii of increasing size and compared the retrieved corrected number counts (`Output') to the underlying source distribution in SIDES (`Input'). 
The results are shown in Fig.~\ref{sim_mask_comp_diff} for Band 6 with radii from 1\,arcsec to 4\,arcsec. 
The lighter shaded region indicates the 1\,$\sigma$ scatter between realisations (i.e. the cosmic variance). 
The darker shaded region indicates the error on the mean over all realisations (i.e. $\sigma/\sqrt{58}$). 
For clarity, the uncertainties are shown only for the Bias line.
While we see a slight convergence, there is no significant overall improvement with increasing radius. 
The excess of the Bias over the Input number counts at the bright end remains consistently at $\sim$6\%.
The deviation, however, is far smaller than the cosmic variance in all flux density bins at any radius. 
Cosmic variance introduces an uncertainty of $\sim$15-20\% in the intermediate flux density ranges and $\sim50\%$ and $\sim90\%$ at the faint and bright end, respectively. 
From a simulation viewpoint, increasing the radius of the mask above 1\,arcsec therefore offers no additional benefit.  

\begin{figure*}
\centering
\includegraphics[width=0.8\linewidth]{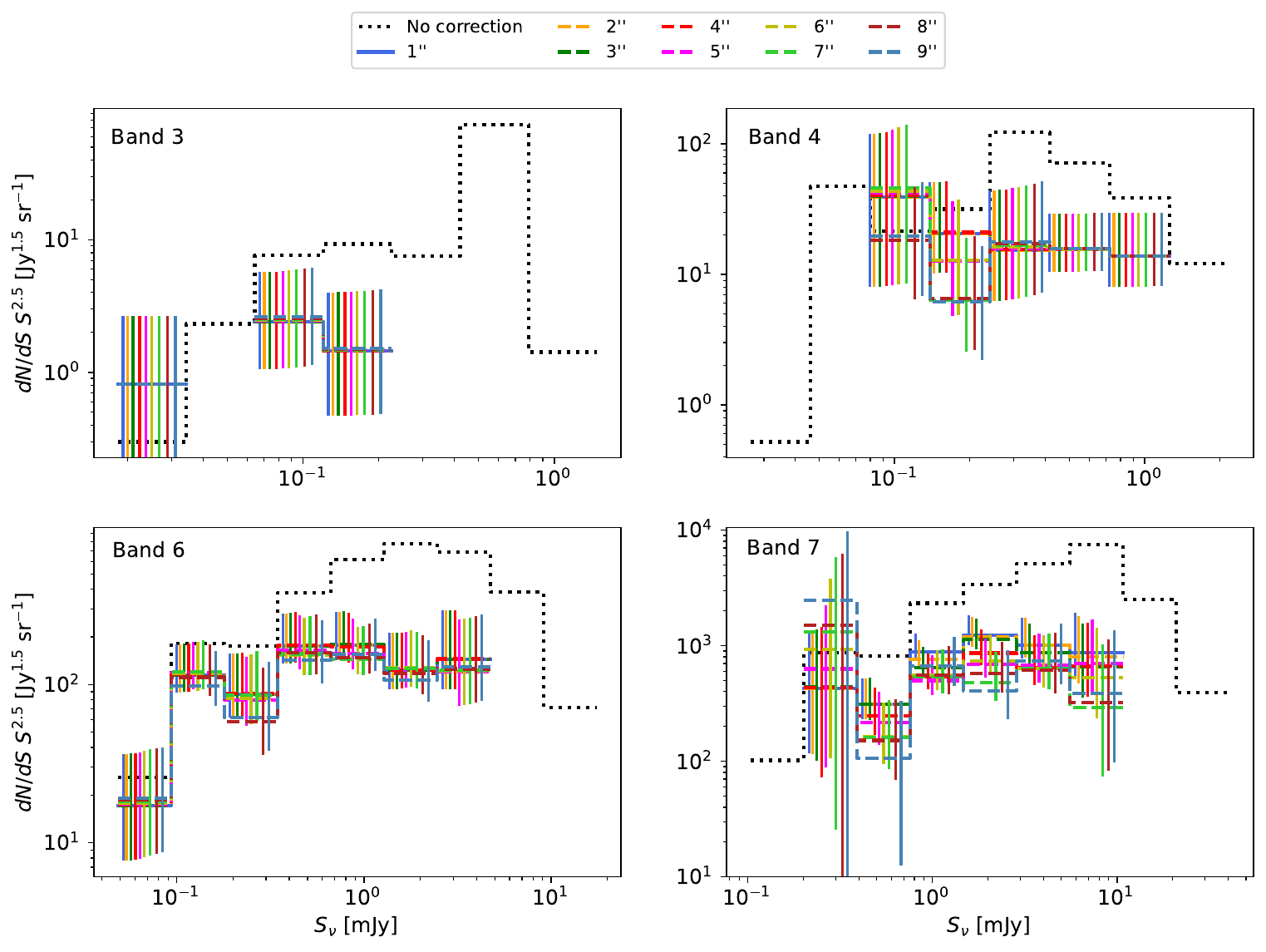}
\caption{Differential continuum number counts computed from the \acosmos database in four ALMA bands (as labelled in the top left corner of each panel) for nine different radii of the mask around the phase centre of single pointings. Uncertainties are shown as vertical coloured lines in the respective bins. The dotted grey line shows the number counts using all data in the \acosmos database and applying no corrections. Error bars are not shown for the uncorrected number counts.}
\label{mask_radii_comp_diff}
\end{figure*}

However, these conclusions are only meaningful if the clustering properties in the simulations are an accurate representation of reality. 
To test the applicability on our \acosmos database, we therefore investigated the influence of varying the mask radius on the inferred number counts. 
Figure~\ref{mask_radii_comp_diff} shows the number counts inferred using our method using different mask radii up to 9\,arcsec as coloured lines. 
In Bands~3, 4, and 6 there is no significant change in the number counts within the uncertainties when increasing the radius beyond 1\,arcsec.
In Band~7, however, there is a slight trend towards lower number counts with increasing radius, but with a converging behaviour. 
For $r\gtrsim4$\,arcsec the number counts are consistent within their uncertainties in all flux density bins. 
This is most likely due to a higher contamination from close-by neighbours of the target source, are not caught as well in this band as in other bands by our redshift weighting method (see Sect.~\ref{subsubsec_pairs}).
Band~7 is the band in which the lowest fraction of serendipitous sources in close angular proximity (i.e. <4\,arcsec) to the target have a redshift counterpart in our ancillary catalogues. 
This reduces our possibility to perform redshift weighting of these sources. 
Thus, while redshift weighting excludes most close neighbours due to clustering in Bands~3, 4, and 6, it is not sufficient in Band~7.
We therefore had to choose a larger exclusion radius in this band.
In brief, we chose for our central mask a radius of 1\,arcsec in Bands~3, 4, and 6, and a radius of 4\,arcsec for Band 7.

\section{Comparing blind surveys and single pointings} \label{appendix_split}

In order to check the validity of our selection, masking and weighting methods on single pointings, we computed the number counts from our \acosmos database for four different cases.
The first is using the entire database, that is, all images and sources, applying no selection, masking or weighting. 
This way we see the full effect of the observational biases in the data. 
The second case is computing the number counts exclusively from blind surveys in our database and the sources detected therein. 
Since blind surveys do not have selection biases by design, no masking or weighting by redshift is necessary to infer accurate number counts. 
The third case is using only single pointings with a central target, applying our masking and weighting method, so we can see the extent to which our method constitutes an improvement.
The fourth case is using blind surveys combined with the selected and masked single pointings, which represents our final results as shown in Fig.~\ref{figure-nc-all}.

\begin{figure*}
\centering
\includegraphics[width=0.8\linewidth]{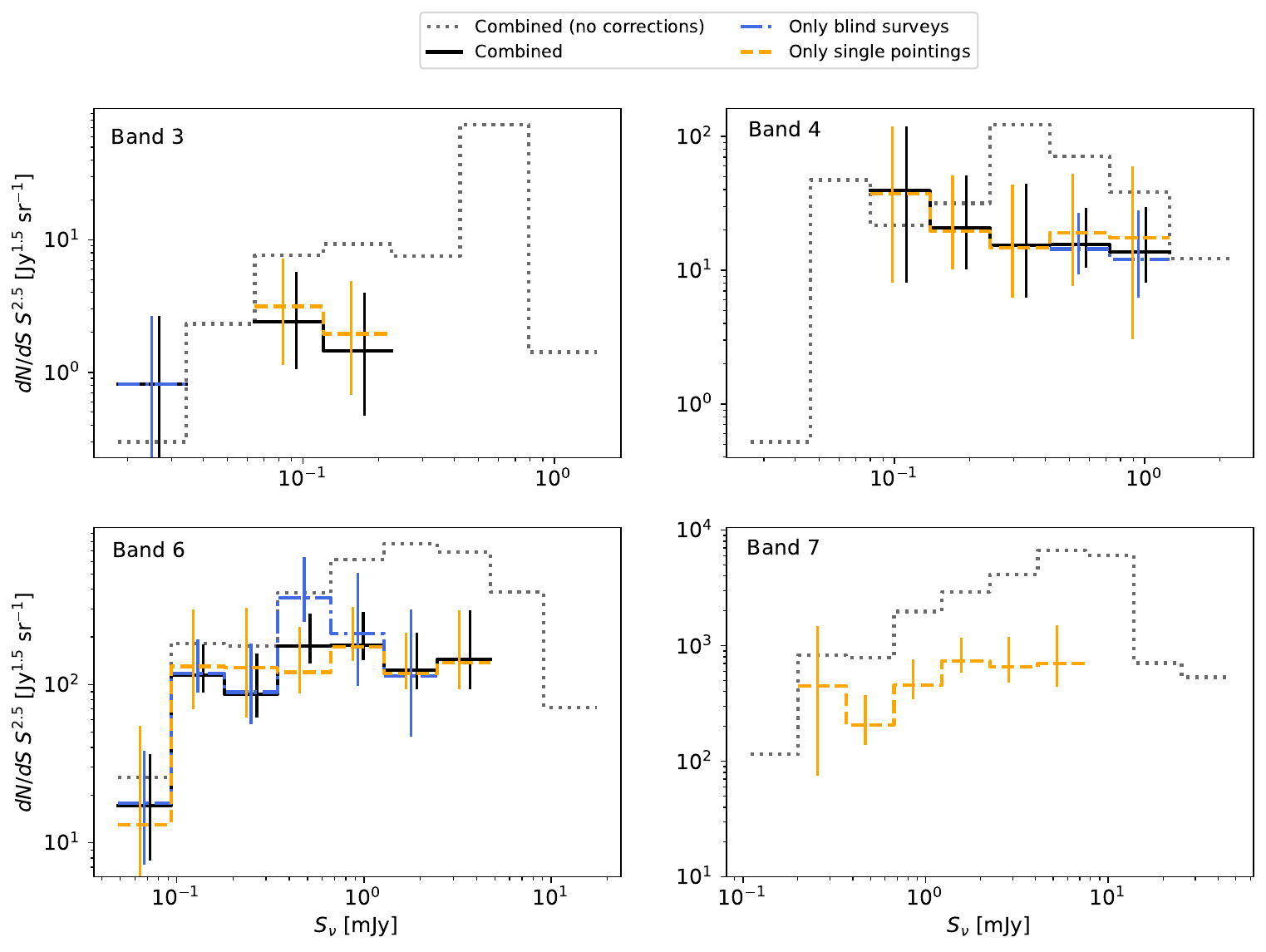}
\caption{Differential continuum number counts computed from the \acosmos database in four ALMA bands (as labelled in the top left corner of each panel). 
The number counts are measured for four different sets of data. 
The grey dotted line indicates use of  all images and blindly extracted sources contained within the entire database, applying no selection or correction; the blue dash-dotted line  is for the use of  images only from ALMA blind surveys and sources found therein; the orange dashed line is for the use of  only precisely targeted single pointings, after masking out a central circular region of 1\,arcsec (4\,arcsec in Band 7) radius and weighting the contribution of each detected source outside the masked region by the proximity of its redshift to that of the respective target (see Sect. \ref{subsec_selection_criteria}); the black solid line is for  both blind surveys and single pointings with targets, after masking the phase centre and weighting sources by redshift. 
As the single pointing line overlaps with other lines in several bins, it has been slightly shifted to improve visibility. In Band 3, it is increased by 5\%, in Band 4 and 6 it is lowered by 5\%. 
Error bars are shown for each bin of all lines except the uncorrected case, in the colour of the respective line. No blind surveys exist in our database for Band 7, and therefore no number counts are shown for that and the combined case.}
\label{versions_comp_diff}
\end{figure*}

The comparison of these four cases is shown in Fig.~\ref{versions_comp_diff}. 
In all bands, there is a significant improvement over the uncorrected number counts, especially towards higher flux densities, which is primarily due to brighter sources being targeted preferentially. 
In Bands~3, 4, and 6, the blind survey, single pointing, and combined number counts are consistent with each other within their range of uncertainties. 
This makes us confident of the validity of our selection and correction method for single pointings to retrieve accurate number counts. 
Blind survey and combined number counts are not shown for Band~7, as there are no blind surveys available for this band within the \acosmos and \agoodss databases.
The only noteworthy difference that can be seen between the blind survey and the combined number counts appears in one flux density bin in Band~6 (i.e. $0.35-0.67$\,mJy). 
While still within the errors, the blind survey count in that bin is by a factor 2 higher than the combined. 
However, this disagreement can entirely be attributed to cosmic variance and small number statistics. 

We conclude from this comparison that our selection and correction method for single pointings is valid, as it yields consistent results with blind surveys. 

\end{document}